\documentclass[useAMS,usenatbib]{mnras}
\usepackage{amssymb}
\usepackage{graphicx}
\usepackage{xspace}
\usepackage{amssymb,latexsym,graphicx,natbib,eufrak,times,amsmath}

\usepackage{ifthen}
\def\draftversion{0} 

\makeatletter
\newcommand\mytoc{%
    \@starttoc{toc}%
}
\makeatother

\ifthenelse{\equal{\draftversion}{0}}{
	\usepackage{xcolor}
	\newcommand{\tmp}{}
	\newenvironment{envcomm}[1]{\renewcommand{\tmp}{#1}\begin{color}[rgb]{0,0.5,0.0}\begin{center}\hrule\vspace{0.5mm}\tmp's COMMENTS\end{center}}{\begin{center}END OF \tmp's COMMENTS\vspace{0.5mm}\hrule\end{center}\end{color}}
	\newenvironment{draft}{\begin{color}[rgb]{0,0.4,0}\begin{center}\hrule\vspace{0.5mm}DRAFT\end{center}}{\begin{center}END OF DRAFT\vspace{0.5mm}\hrule\end{center}\end{color}}
	\newcommand{\comcomm}[2]{\begin{color}[rgb]{0,0.5,0.0}\ $\bullet$ \textbf{#1:} #2 $\bullet$\ \end{color}}
	\newcommand{\revend}[1]{\par\begin{color}[rgb]{0,0.4,0}\begin{center}\hrule\vspace{0.5mm}END OF #1's REVISIONS\vspace{0.5mm}\hrule\end{center}\end{color}\par}
	\newcommand{\todo}[1]{\begin{color}{red}\ $\bullet$ \textbf{To do: }#1 $\bullet$\ \end{color}}
	
	\newcommand{\del}[1]{\begin{color}[rgb]{0,0.5,0.0}\ $\bullet$ \textbf{Deleted: }#1 $\bullet$\ \end{color}}
	\newcommand{\sk}[1]{\begin{color}[rgb]{0.6,0,0.6}#1\end{color}}
	\newcommand{\toc}{\par\begin{color}[rgb]{0.6,0,0.6}\begin{center}\hrule\vspace{0.5mm}\begingroup\small\let\cleardoublepage\relax\let\clearpage\relax\mytoc\endgroup\vspace{0.5mm}\hrule\end{center}\end{color}\par}
	}{
	\newsavebox{\trashcan}
	\newenvironment{envcomm}[1]{\begin{lrbox}{\trashcan}\begin{minipage}{\columnwidth}}{\end{minipage}\end{lrbox}}
	
	\newcommand{\comcomm}[2]{}
	\newcommand{\revend}[1]{}
	\newcommand{\todo}[1]{}
	
	\newcommand{\del}[1]{}
	\newcommand{\sk}[1]{}
	\newcommand{\toc}{}
	}

\long\def\symbolfootnote[#1]#2{\begingroup%
\def\thefootnote{\fnsymbol{footnote}}\footnote[#1]{#2}\endgroup} 

\newcommand{\eqn}[2][]{Equation#1~\ref{eqn:#2}} 
\newcommand{\fig}[2][]{Figure#1~\ref{fig:#2}}

\newcommand{\sect}[2][]{Section#1~\ref{sec:#2}}
\newcommand{\app}[2][]{Appendix#1~\ref{sec:#2}}
\renewcommand{\eqn}[2][]{equation#1~(\ref{eqn:#2})}
\renewcommand{\fig}[2][]{Fig#1.~\ref{fig:#2}}

\newcommand{\mh}{\ensuremath{\textrm{\,--\,}}}
\newcommand{\bb}[1]{\ifmmode \mbox{\boldmath $ #1$} \else  \mbox{\boldmath $#1$} \fi}

\newcommand{\ave}[1]{\langle #1 \rangle}
\newcommand{\U}[1]{\ensuremath{\mathrm{~#1}}}
\newcommand{\e}[1]{\ensuremath{\times 10^{#1}}}
\newcommand{\yr}{\U{yr}}
\newcommand{\Myr}{\U{Myr}}
\newcommand{\Gyr}{\U{Gyr}}
\newcommand{\pc}{\U{pc}}
\newcommand{\kpc}{\U{kpc}}
\newcommand{\Mpc}{\U{Mpc}}
\newcommand{\msun}{\U{M}_{\odot}}
\newcommand{\Msun}{\msun}
\newcommand{\Msunyr}{\Msun\yr^{-1}}
\newcommand{\Zsun}{\U{Z}_{\odot}}
\newcommand{\cc}{\U{cm^{-3}}}
\newcommand{\K}{\U{K}}

\newcommand{\kms}{\U{km\ s^{-1}}}
\newcommand{\erg}{\U{erg}}

\newcommand{\ramses}{{\small RAMSES}\xspace}

\newcommand{\hop}{{\small HOP}\xspace}
\newcommand{\starburst}{{\small STARBURST99}\xspace}
\newcommand{\music}{{\small MUSIC}\xspace}
\newcommand{\lambdasun}{\bar{\lambda}_{\odot}}

\title[The origin of the Milky Way globular clusters]{The origin of the Milky Way globular clusters}

\author[Renaud, Agertz \& Gieles] {Florent~Renaud$^1$\thanks{f.renaud@surrey.ac.uk}, Oscar~Agertz$^{1,2}$ and Mark Gieles$^1$\\
$^1$Department of Physics, University of Surrey, Guildford, GU2 7XH, UK\\
$^2$Lund Observatory, Department of Astronomy and Theoretical Physics, Box 43, SE-22100, Lund, Sweden
}

\date{Accepted 2016 November 14. Received 2016 November 13; in original form 2016 October 3}

\begin{document}
\maketitle


\begin{abstract}
We present a cosmological zoom-in simulation of a Milky Way-like galaxy used to explore the formation and evolution of star clusters. We investigate in particular the origin of the bimodality observed in the colour and metallicity of globular clusters, and the environmental evolution through cosmic times in the form of tidal tensors. Our results self-consistently confirm previous findings that the blue, metal-poor clusters form in satellite galaxies which are accreted onto the Milky Way, while the red, metal-rich clusters form mostly in situ or, to a lower extent in massive, self-enriched galaxies merging with the Milky Way. By monitoring the tidal fields these populations experience, we find that clusters formed in situ (generally centrally concentrated) feel significantly stronger tides than the accreted ones, both in the present-day, and when averaged over their entire life. Furthermore, we note that the tidal field experienced by Milky Way clusters is significantly weaker in the past than at present-day, confirming that it is unlikely that a power-law cluster initial mass function like that of young massive clusters, is transformed into the observed peaked distribution in the Milky Way with relaxation-driven evaporation in a tidal field.
\end{abstract}

\section{Introduction}

Globular clusters are among the oldest astrophysical objects. They form in the early Universe in highest density peaks \citep[e.g.][]{Diemand2005, Boley2009} and because of their high density, they survive tidal harassment from their host(s). Hence, they witness most of the formation and evolution processes of galaxies, and can be used to probe them \citep{Brodie2006}. The globular clusters populations of most (it not all) massive galaxies display a colour bimodality: the so-called blue and red clusters \citep[see e.g.][and references therein]{Zinn1985, Gebhardt1999, Larsen2001, Peng2006}. Due to their extended distribution in space, blue clusters used to be referred to as halo clusters. They are metal-poor (with the distribution peaking at [Fe/H] $\approx -1.5$ for the Milky Way, \citealt{Harris1996}, but this value varies from galaxy to galaxy) and show no sign of rotation as a population, contrary to the red clusters which are more metal-rich (peak at [Fe/H] $\approx -0.5$ in the Milky Way), more spatially concentrated and which are rotating with the galaxy \citep[see a review in][]{Brodie2006}. 

Such bimodality suggests two formation mechanisms for globular clusters. For instance, it has been proposed that blue clusters form in galaxies in the early Universe that merge later. In such (wet) merger process, starbursts would generate the red population \citep{Schweizer1987, Ashman1992}. The resulting massive galaxy would then exhibit its own red population together with the blue clusters of its progenitors. This scenario is supported by the observed ability of wet mergers to produce young massive clusters \citep[see also \citealt{Renaud2014b, Renaud2015a} for a theoretical perspective]{Whitmore1995, Mengel2005, Bastian2009, Herrera2011} which could share the same physical properties as (present-day observed) globular clusters when they formed. However this scenario faces issues, mainly in matching the specific frequency of globular clusters, in particular in massive elliptical galaxies \citep[e.g.][]{McLaughlin1994}. \citet{Forbes1997} propose instead that blue globulars form when the proto-galaxy itself collapses, in a metal-poor and turbulence media. The red population would form later, once the galactic disc has settled. The formation of globular clusters would then be a multiphase process, with the first phase being interrupted possibly by cosmic reionisation \citep{Beasley2002}.

\citet{Kravtsov2005} and \citet{Li2014} advocate that major mergers are at the origin of both sub-populations: blue clusters form during early mergers ($z > 4$) while the red ones appear in mergers at lower redshifts (even after $z=1$). Although this scenario combined with star formation enhancement in mergers seems appropriate in dense galactic environment leading to the assembly of massive elliptical galaxies, like in the Virgo Cluster as tested by \citet{Li2014}, it does not apply to Milky-Way like systems where no recent major merger took place \citep{Wyse2001, Deason2013, Ruchti2014, Ruchti2015}. Finally, \citet{Cote1998} argue that red clusters form in situ while the blue ones are accreted, either via merging satellite galaxies, or by tidal capture of the clusters themselves \citep[see also][]{Tonini2013}.

Because of the age-metallicity degeneracy \citep[e.g.][]{Worthey1994}, determining the physical origins of the two populations is an observational challenge \citep[see][]{Marin2009, Leaman2013, Forbes2015}, and the numerical approach has been followed by several pioneer works to shed more light on this topic. There, the difficulty consists in the wide range of scales (space and time) spanning both the cosmological context of galaxy formation and evolution, and the internal physics of clusters. For this reason, all previous studies focus on a specific aspect of the problem. For instance, following \citet{Kravtsov2005}, \citet{Li2016} use parsec-resolution cosmological simulations to explore the physics of globular cluster formation, in particular the shape of the cluster initial mass function, leaving aside the environment-dependent cluster evolution.

Modelling the evolution of clusters requires to resolve their internal physics driven by stellar evolution and interactions between individual stars (and/or binary and multitple systems, see e.g. \citealt{Heggie2003}), which is still out of reach of large scale galactic and cosmological simulations. However, several works develop and use coupling methods in which the cosmological environment is (partially) accounted for in dedicated $N$-body star-by-star simulations of clusters \citep{Fujii2007, Renaud2011, Renaud2015b}. Studies consider either the full cosmological context \citep{Rieder2013}, or focus on specific aspects of galaxy evolution like major mergers \citep{Renaud2013a}, accretion of satellites \citep{Miholics2014, Bianchini2015} or the secular growth of the host galaxy’s dark matter halo \citep{Renaud2015c}. All these works point out the mild effects of the cosmological context and the galactic events on the short time-scale evolution of clusters ($\lesssim$ Gyr, as opposed to local perturbations like encounters with molecular clouds \citealt{Elmegreen2010, Gieles2016}). However, these evolutions modify the context itself (e.g. the galactic mass distribution) and/or the orbits of the clusters, which influence their evolution over longer time-scales ($\gtrsim$ Gyr). Connecting the theory on cluster formation and their response to the evolution of their environment over several Gyr to the observation of globular clusters is thus an on-going process in the understanding of origin of galaxies and of their stellar populations.

Using a cosmological zoom-in simulation of the formation of a Milky Way-like galaxy, we explore both the formation and the evolution of star clusters, tracking their orbits and the tidal field they experience since their formation, in the Milky Way and its progenitors. We particularly focus on the bimodality of globulars retrieved in our simulation and explore the differences between the two sub-populations along their evolutions to propose a consistent scenario of their origins.

\section{Methodology}
\label{sec:method}

We carry out a cosmological hydrodynamic+$N$-body zoom-in simulation of a Milky Way mass galaxy using the adaptive mesh refinement (AMR) code \ramses \citep{Teyssier2002}, assuming a flat $\Lambda$-cold dark matter cosmology with $H_0 = 70.2 \U{km\ s^{-1}\ Mpc^{-1}}$, $\Omega_{\rm m} = 0.272$, $\Omega_\Lambda = 0.728$, and $\Omega_{\rm b}= 0.0455$. A dark matter only simulation was performed beforehand using a simulation cube of size $85 \Mpc$. At $z=0$, a halo of $R_{200m} = 334 \kpc$ (radius within which the mass density is 200 times the mean matter density) and $M_{200m} = 1.3\e{12} \Msun$ was selected for re-simulation at high resolution. Particles within $5 R_{200m}$ at $z=0$ were traced back to $z=100$, and the Lagrangian region they defined was regenerated at high resolution, still embedded within the full lower-resolution volume, using the \music code \citep{Hahn2011}. This is the same initial conditions as the ``m12i'' halo from \citet{Hopkins2014} and \citet{Wetzel2016}. 

The dark matter particle mass in the high resolution region is $2.1 \e{6} \Msun$ and the adaptive mesh is allowed to refine if a cell contains more than eight dark matter particles. This allows the local force softening to closely match the local mean inter-particle separation, which suppresses discreteness effects \citep[e.g.,][]{Romeo2008}. A similar criterion is employed for the baryonic component, where the maximum refinement level is set to allow for a mean constant physical resolution of $218 \pc$.

The adopted star formation and feedback physics is presented in \citet{Agertz2013} and \citet{Agertz2015, Agertz2016}. Briefly, star formation proceeds in gas denser than $1 \cc$. At our working resolution, choosing a threshold better representative of molecular ($10^2 \cc$, \citealt{Krumholz2009a}) or self-gravitating gas ($10^3 \cc$, \citealt{Renaud2013b}) would not provide a good match of the star formation rate, in particular in the low-mass haloes. Above this threshold, our star formation prescription is based on the local abundance of molecular hydrogen following \citet{Krumholz2009a}, and each formed stellar particle is treated as a single-age stellar population with a \citet{Chabrier2003} initial mass function. We account for injection of energy, momentum, mass and heavy elements over time via type II and type Ia supernovae (SNe), stellar winds and radiation pressure (allowing for both single scattering and multiple scattering events on dust) on the surrounding gas. Each mechanism depends on the stellar age, mass and gas/stellar metallicity, calibrated on the stellar evolution code \starburst \citep{Leitherer1999}. Feedback is done at the appropriate times, taking into account the lifetime of stars of different masses in a stellar population through the metallicity dependent age-mass relation of \citet{Raiteri1996}. 

Furthermore, we adopt the SN momentum injection model recently suggested by \citet[see also \citealt{Martizzi2015, Gatto2015, Simpson2015}]{Kim2015}. Here we consider a SN explosion to be resolved when the cooling radius\footnote{The cooling radius in gas of density $n$ and metallicity $Z$ scales as $\approx 30 (n/1\cc)^{-0.43} (Z/\Zsun + 0.01)^{-0.18} \pc$ for a supernova explosion with energy $E_{\rm SN}=10^{51}$ erg \citep[e.g.][]{Cioffi1988, Thornton1998}.} is resolved by at least three grid cells. In this case the explosion is initialized in the energy conserving phase by injecting the relevant energy ($10^{51} \erg$ per SN) into the nearest grid cell. If this criterion is not fulfilled, the SN is initialized in its momentum conserving phase, i.e. the total momentum generated during the energy conserving Sedov-Taylor phase is injected into to the 26 cells surrounding a star particle. It can be shown \citep[e.g.][]{Blondin1998, Kim2015} that at this time, the momentum of the expanding shell is approximately $2.6\e{5} (E_{\rm SN}/10^{51}\erg)^{16/17} (n/1\cc)^{-2/17} \U{\Msun\kms}$.
 
Heavy elements (metals) injected by supernovae and winds are advected as a passive scalar and are incorporated self-consistently in the cooling and heating routine assuming solar composition. The code accounts for metallicity dependent cooling by using tabulated cooling functions of \citet{Sutherland1993} for gas temperatures $10^{4\mh8.5} \K$, and rates from \citet{Rosen1995} for cooling down to lower temperatures. Heating from the ultraviolet background radiation is accounted for by using the model of \citet{Haardt1996}, assuming a reionization redshift of $z=8.5$. We follow \citet{Agertz2009b} and adopt an initial metallicity of $Z = 10^{-3} \Zsun$ in the high-resolution zoom-in region in order to account for enrichment from unresolved population III star formation \citep[e.g.][]{Wise2012}. 

By accounting for the above processes, and allowing for star formation to be locally efficient, \citet{Agertz2015, Agertz2016} demonstrated that feedback regulation resulted in a galaxy corresponding to a realistic late-type galaxy that matches the evolution of basic properties of late-type galaxies such as stellar mass, disc size, the presence of a thin and thick stellar disc, stellar and gas surface density profiles, the Kennicutt-Schmidt relation, the stellar mass-gas metallicity relation (and its evolution), and a specific angular momentum typical of spiral galaxies of the Milky Way mass (stellar mass of $\approx 5\e{10} \msun$).

The simulation is run until it reaches $z=0.5$ (i.e. a lookback time of about $5 \Gyr$), i.e. long after the last major merger event which occurs at $z\approx 2$ (see \sect{galaxyformation}). The simulation has been performed on the \emph{Curie} supercomputer hosted at the \emph{Tr\`es Grand Centre de Calcul} (TGCC).

We identify galaxies using the friend-of-friend algorithm \hop \citep{Eisenstein1998} on the stellar component only, with the parameters $10^5, 10^4, 10^2 \Msun\kpc^{-3}$ for the peak, saddle and outer densities respectively, chosen to agree with a visual examination of stellar density maps. For convenience, hereafter we refer as the Milky Way to the most massive progenitor of the most massive galaxy in our simulated volume at $z=0.5$. We identify the plane of its disc using the total angular momentum vectors of all stars within $10 \kpc$ from its centre of mass, and use the prime notation for quantities projected into the cylindrical coordinate system defined by the plane of the disc and its spin axis. We defined stars formed in situ those which have formed in the most massive progenitor of the Milky Way, and are still in the progenitor of the Galaxy at a given time, thus excluding stars that formed in the Milky Way but have been ejected later on, without being re-accreted. Accreted stars are all other stars in the Milky Way. 

The originality of our study consists in combining both the formation and the evolution of star clusters. From the formation point of view, details on the structure of the interstellar medium (ISM) and the propagation of stellar feedback at the scale of molecular cloud are not resolved in our simulation, which thus affects the mass and size of our stellar objects. The mass of our stellar particles ($\sim 10^4 \Msun$) forbids the direct identification of star clusters. Not being able to resolve stellar systems less massive then several times this mass resolution, and with the gravitation affected by softening at the resolution of the AMR grid which introduces artefacts on the boundness of aggregates of such particles into ``clusters'', we do not seek an identification of star clusters and rather use our stellar particles as tracers of both the conditions and epochs of their formation and of their orbits. Therefore, we refer to them as ``star cluster candidates'' in the following. Among this group, we further qualify the subset of particles formed before a lookback time of $10 \Gyr$ ($z \gtrsim 1.8$) as ``globular cluster candidates'', in line with observational definitions \citep[e.g.][]{Portegies2010}. From the evolution perspective, our simulation does not capture the internal dynamics of the clusters, which is known to be an important driver of their evolution \citep{Henon1961, Gieles2011b}. A full coupling of the internal and external physics of star clusters is still out of reach due to technical limitations, but previous works provide an estimate of the gravitational (tidal) effect of a non-regular galactic or cosmological environment on star clusters \citep[e.g.][]{Renaud2011, Rieder2013}. We adopt this approach by computing the tidal tensor along the orbits of our cluster candidates (See \app{tensor} for details). The tensors are evaluated as the first order finite differences of the gravitational acceleration on the AMR cell containing the particle, as in \citet{Renaud2014b}. By doing this at the (local) resolution of the grid (instead of interpolating the acceleration at smaller scales), we ensure that the gravitational information we extract is not strongly affected by the artificial softening of the gravitational acceleration (see also \citealt{Renaud2010a}, Chapter 1 for details). Following all stellar particles to trace their formation, orbit and tidal history would be numerically costly. In particular, the computation of the tidal tensors and its output for all particles at a high frequency (about every Myr) would significantly slow down the simulation. Instead, we defined a subset of about 15\,000 stellar particles in the region of the Milky Way progenitor and extract tidal data for the selected subset only. This is done at every coarse timestep, i.e. $\approx 0.5 \mh 5 \Myr$. We have checked that the subset is representative of the environment of the Milky Way by ensuring it follows the same distributions in space, age and metallicity.

\section{Results}

\subsection{Galaxy and star formation}
\label{sec:galaxyformation}

\begin{figure*}
\includegraphics{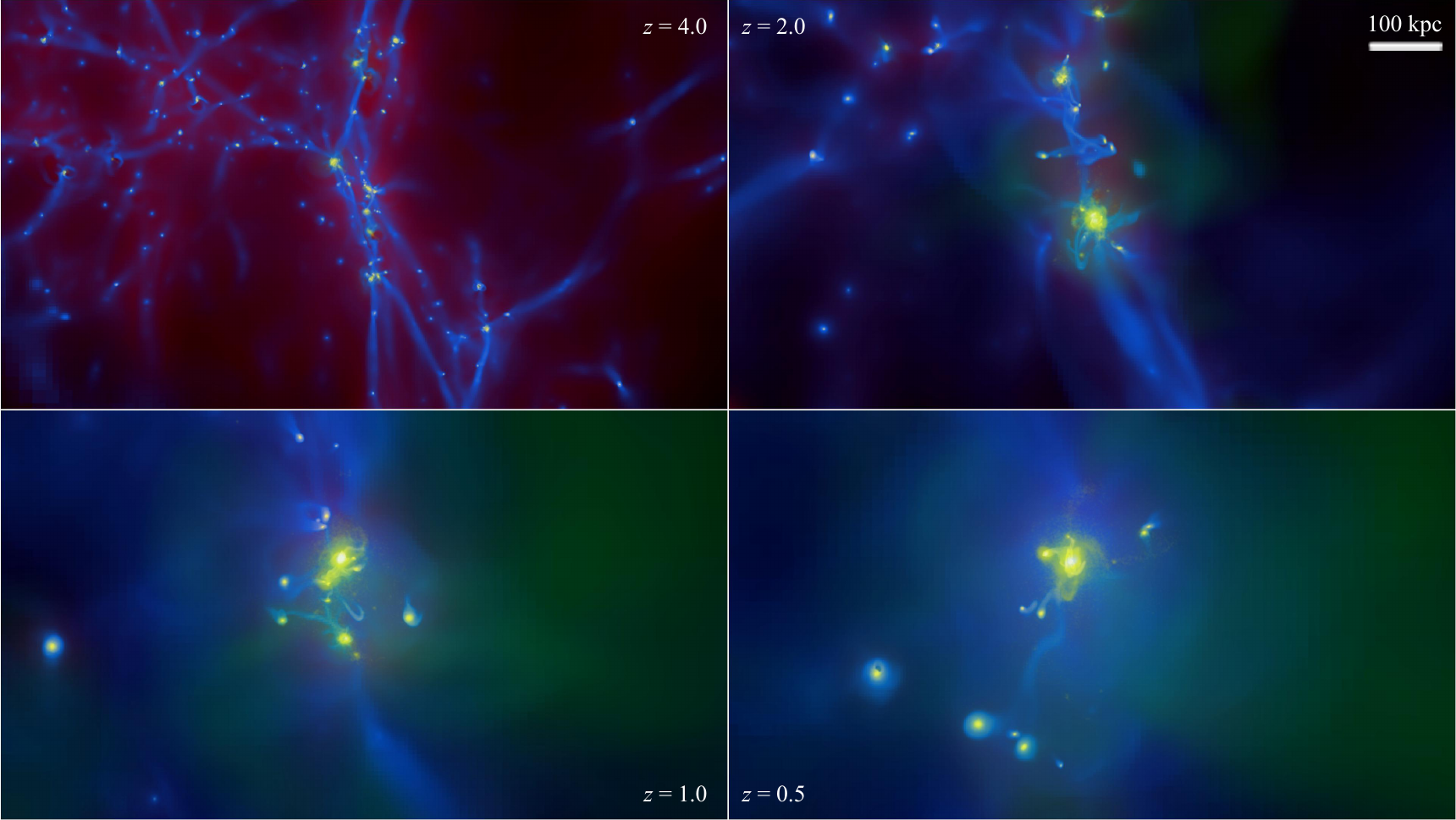}
\caption{Morphology of simulated neighbourhood of the Milky Way at four epochs, showing the gas density (blue), dark matter (red), stars (yellow) and iron density (green).}
\label{fig:morpho}
\end{figure*}

\begin{figure*}
\includegraphics{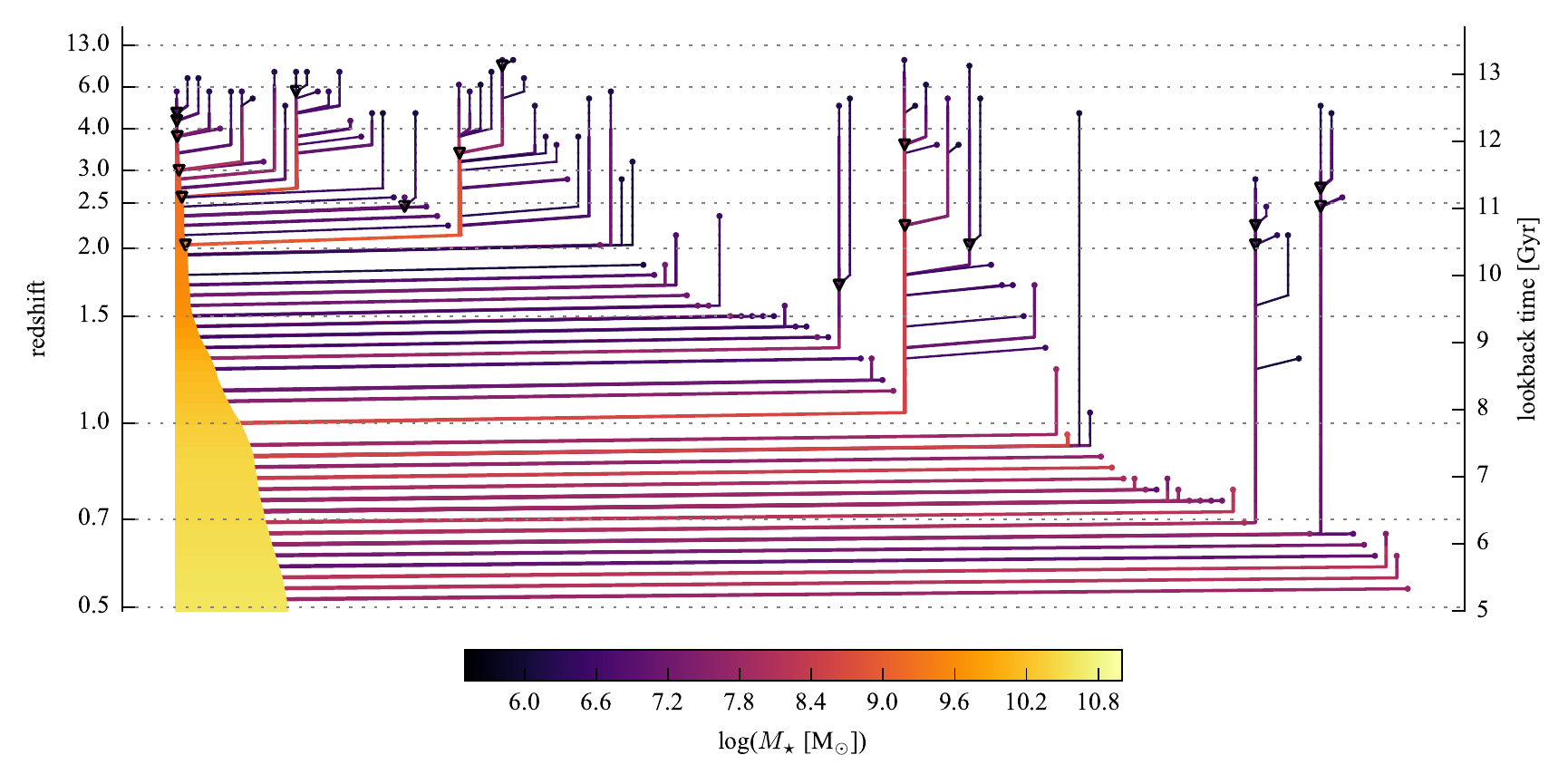}
\caption{Stellar merger tree of Milky Way. Dots mark the earliest detection of a stellar clump. Colours and line widths indicate the stellar mass of the galaxy. Triangles mark major mergers, i.e. mergers with a mass ratio greater than 1:10. The tree is computed at the output frequency of the simulation, i.e. $\approx 150 \Myr$.} 
\label{fig:mergertree}
\end{figure*}

\fig{morpho} shows the morphology of the Milky Way neighbourhood at four epochs along the course of galaxy formation\footnote{Movies are available at \\{\tt http://personal.ph.surrey.ac.uk/$\sim$fr0005/movies.php}}. As a complement, \fig{mergertree} shows the merger tree of the Milky Way, based on the stellar component only. At $z=4$, the intergalactic filaments can still be clearly identified. The main galaxies are found at their intersection \citep[as expected, see e.g.][]{Zeldovich1982} and are fuelled with gas and smaller galaxies. At $z \approx 3.5$, the main two gaseous filaments (almost vertical thin blue structures in the top-left panel of \fig{morpho}) collide, further fuelling galaxies with gas and producing shocks that trigger an episode of star formation. At $z \approx 2$, our Milky Way experiences its last major merger (i.e. with a mass ratio higher than 1:10), but is still surrounded by numbers of small satellites. The starbursts and stirring of the ISM following (major and minor) mergers allows the injection of metals by stellar feedback and their propagation in the interstellar and intergalactic media. At $z=1$, most of the growth of the Milky Way is done, but several small satellites remain in the vicinity of the Galaxy. Most of them are accreted before $z=0.5$. At $z=0.5$ (which corresponds to the end of our simulation), the simulated Milky Way has a stellar mass of $4.1 \e{10} \msun$.

\begin{figure}
\includegraphics{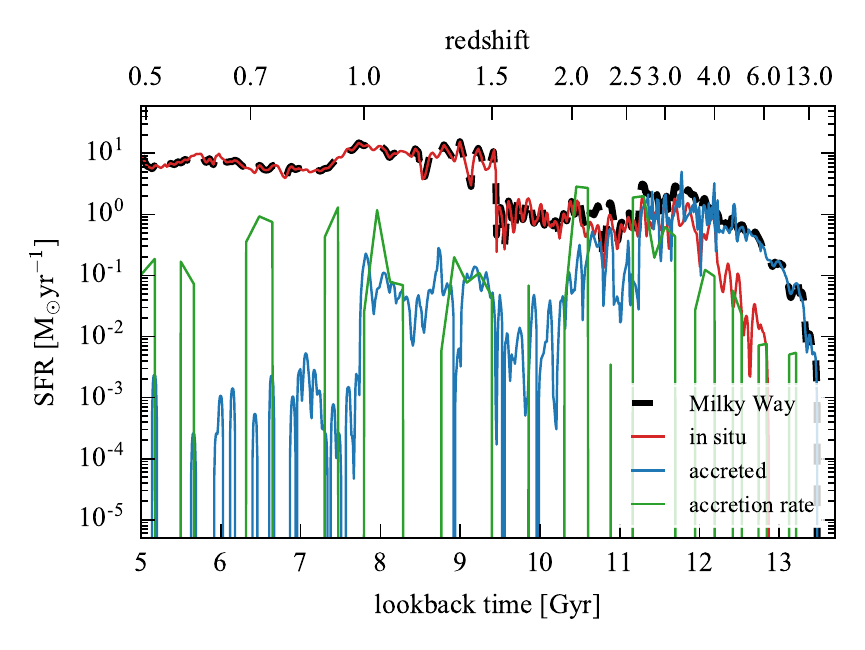}
\caption{Star formation history (computed as a mass-weighted histogram of formation times of the stars detected in the Milky Way progenitor at $z=0.5$, black), distinguishing the stars formed in situ (red) from those accreted (blue). The green curve shows the stellar accretion rate.}
\label{fig:sfr}
\end{figure}

\fig{sfr} displays the star formation history (SFH) of the Milky Way and tells apart the in situ and accreted populations\footnote{The jump in star formation rate (SFR) at $z\approx 1.5$ originates from the activation of an additional refinement level: refinement is artificially blocked when the gas mass in the AMR cells is too low compared to the mass of the stellar particle. This happens at high redshift and is suddenly deactivated, as already noted by \citet{Agertz2011}. In reality (and at higher resolution), this jump would be replaced by a slower increase of the SFR, starting at higher redshift and reaching the maximum value at the epoch of the jump we detect. The SFR and other quantities presented in this paper are thus altered by this artefact but only a few $100 \Myr$ \emph{before} the jump, which does not affect our conclusions.}. The SFH starts by being dominated by the accreted population, indicating that the Milky Way progenitor assembles mostly through the merging of smaller galaxies. Until $z \approx 2$, maxima in the stellar accretion rate (corresponding to the discrete events of major mergers) are of the same order as the SFR, i.e. $\approx 1\mh3 \Msunyr$. At later stages, the accretion rate peaks at comparable values, which indicates that the satellite galaxies merging with the Milky Way remain of comparable mass throughout the evolution. However, because the Milky Way itself keeps growing from accretion and in situ star formation, the late mergers become minor and have less influence on the mass budget and dynamics of the main galaxy. We note that, according to our definition of formation before a lookback time of $10 \Gyr$, globular cluster candidates form during this merger-dominated phase, in contrast with the more quiet period of secular evolution taking place after. The SFR reaches it maximum between $z=1$ and $2$, in line with observational data of galaxies in the mass range of the Milky Way \citep[e.g.][]{Leitner2012, vanDokkum2013}.

\begin{figure}
\includegraphics{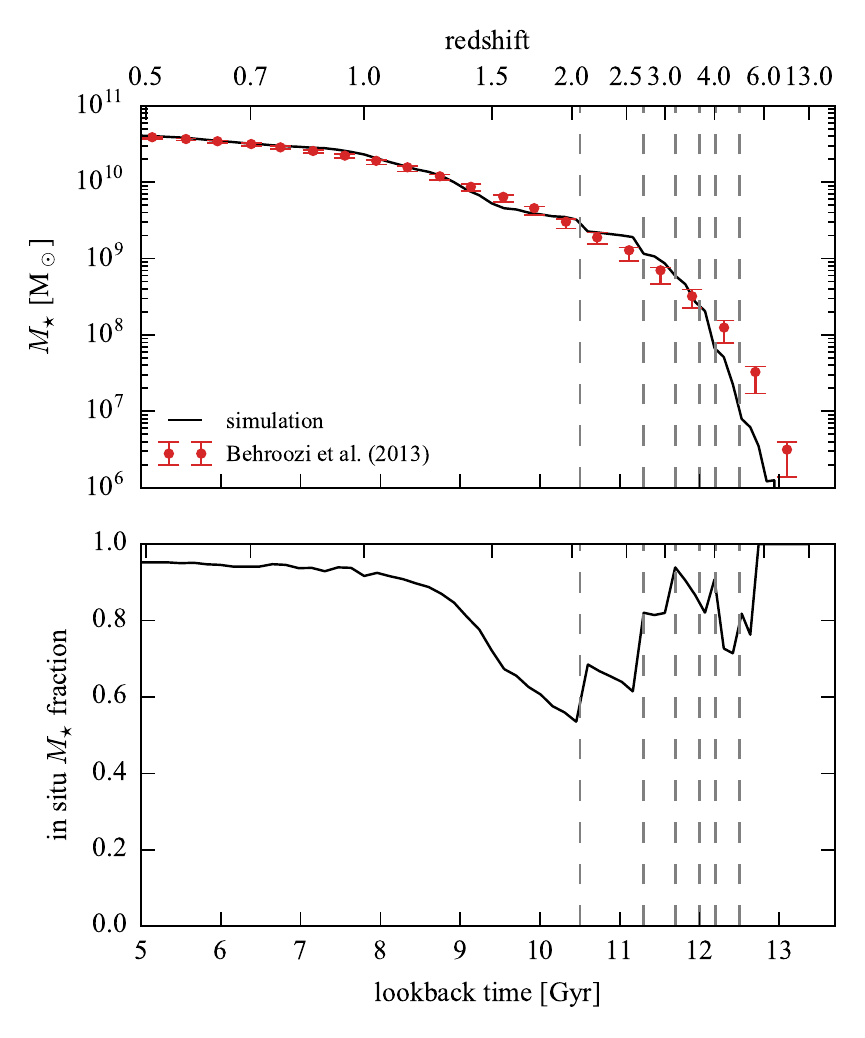}
\caption{Top: evolution of the stellar mass of the simulated Milky Way, compared to the semi-empirical model of \citet{Behroozi2013}. Bottom: stellar mass fraction of the stars formed in situ compared to the total Milky Way progenitor (i.e. the sum of in situ and accreted populations). Vertical dashed lines indicates the major mergers, as identified in \fig{mergertree}.}
\label{fig:isfrac}
\end{figure}

\fig{isfrac} shows the evolution of the stellar mass, compared to the model of \citet{Behroozi2013}\footnote{scaled for a stellar mass of $4.9 \e{10} \msun$ and a halo mass of $10^{12} \Msun$ at $z=0$, which corresponds to the same stellar mass to halo mass as that of the Milky Way ($1.3 \times 10^{12} \msun$ and $6.4 \e{10} \Msun$, \citealt{McMillan2011}).}. Except at very high redshift ($z \gtrsim 4 \mh 5$) when it is sensitive to the (poorly constrained) merger history, the build up of the simulated galaxy corresponds well to the model data.

The bottom panel of the \fig{isfrac} displays the fraction of stars formed in situ. Rapid drops in this fraction denote a ``dillution'' of the in situ stars following the accretion of an external galaxy by the Milky Way, while its slow increase indicates the secular conversion of gas into stars inside the galaxy. We note that, as expected from the merger tree (\fig{mergertree}), rapid changes associated with major mergers dominate the evolution before $z \approx 2$, while a more regular behaviour is found later, in line with the SFH of the two sub-populations (\fig{sfr}). At $z=0.5$, $95\%$ of the stars in the Milky Way have formed in situ, which is compatible with the model of \citet{Behroozi2013}. This is however higher than the value of $75 \%$ (at $z=0$) found in the ``Eris'' simulation \citep{Guedes2011} by \citet{Pillepich2015}, likely because of their intense ex situ SFR, and not because of the galactic interaction history which is qualitatively comparable to that of our model (see their figure 3).

\begin{figure}
\includegraphics{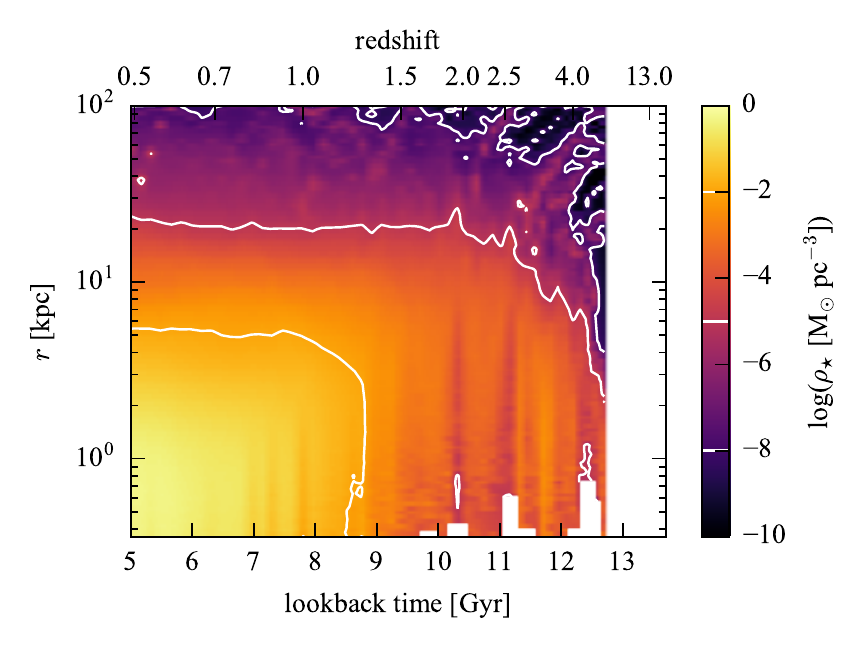}
\caption{Evolution of the radial stellar density profile, including all stars (Milky Way and other galaxies). Contours roughly separate the regions of the disc, the outer galaxy and its satellites.}
\label{fig:profile}
\end{figure}

\fig{profile} displays the evolution of the stellar density profile of the galaxy. Incoming galaxies are visible as oblique patterns in the outer regions (highlighted by the contours), and redistribution of the mass in the inner regions during the major merger events ($z \gtrsim 2$). The main growth phase by accretion ends at $z\approx 2$. Later the stellar component gets denser (i.e. more massive but not more extended) because of in situ star formation. This figure suggests the formation of the disc at $z \approx 1.2 \mh 1.5$.

In short, the formation history, merger tree and build-up of our Milky Way are in-line with literature on the formation of the Galaxy \citep[see e.g.][and references therein]{Freeman2002}. In particular, the absence of major mergers after $z \approx 2$ agrees with predictions from the observed structures of the disc(s) and the lack of tidal signatures in the real Galaxy \citep{Wyse2001, Brodie2006}.

\subsection{Stellar distributions, metallicities and kinematics}
\label{sec:distrib}

\begin{figure}
\includegraphics[scale=0.95]{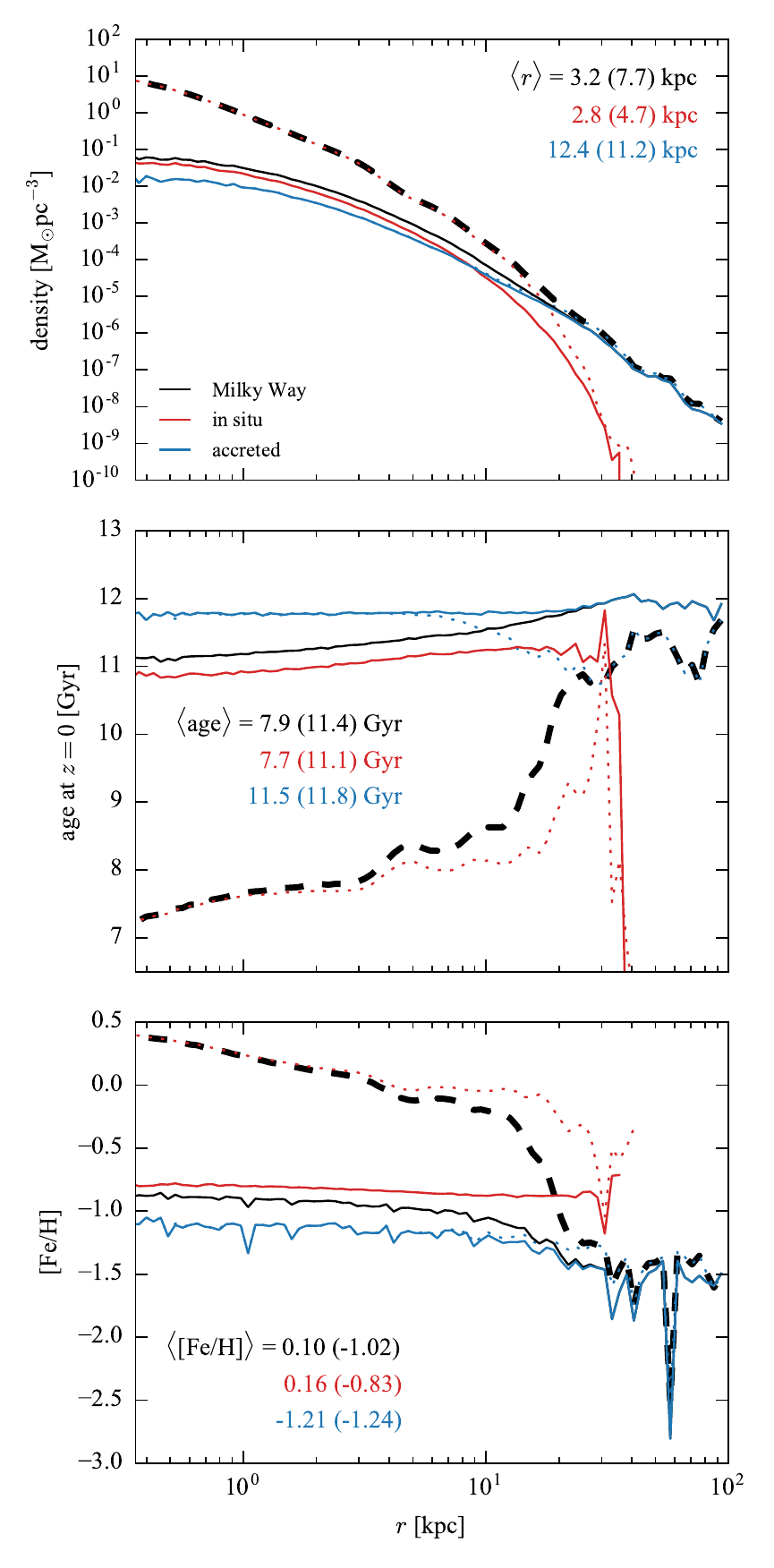}
\caption{Profiles of the stellar mass (top), the age (middle) and the metallicity (bottom) of the Milky Way stars (dotted lines), distinguishing those formed in situ from those accreted. Solid lines represent globular cluster candidates only (i.e. formed before a lookback time of $10 \Gyr$). The thick black dashed line represent all our cluster candidates, i.e. all the stellar particles. Numbers indicate average values for the entire (sub-)populations, and for globular clusters only in parentheses. The dashed black line in the top-panel corresponds to the left-most vertical slice of \fig{profile}, minus the contribution of other galaxies than the Milky Way.}
\label{fig:distrib}
\end{figure}

\begin{figure}
\includegraphics{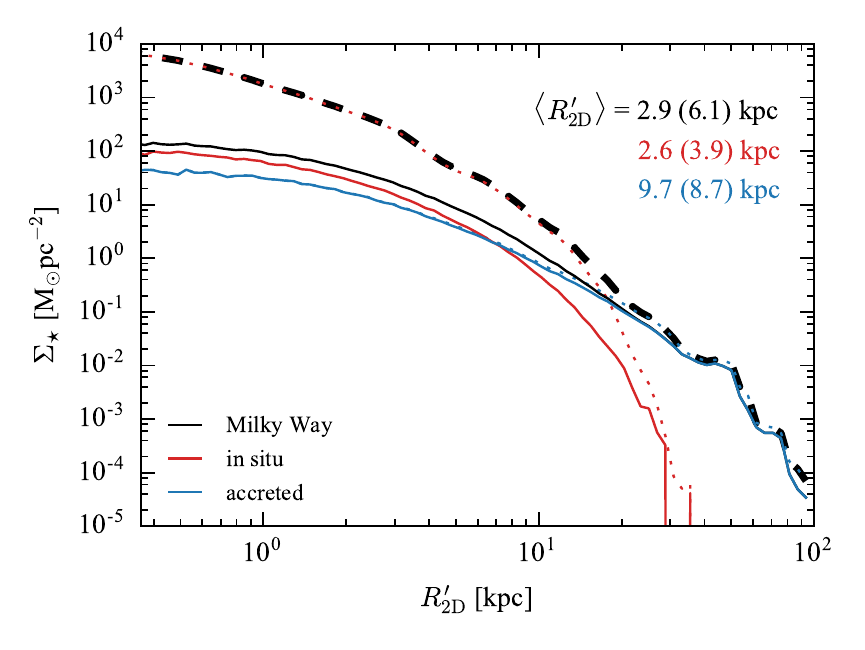}
\includegraphics{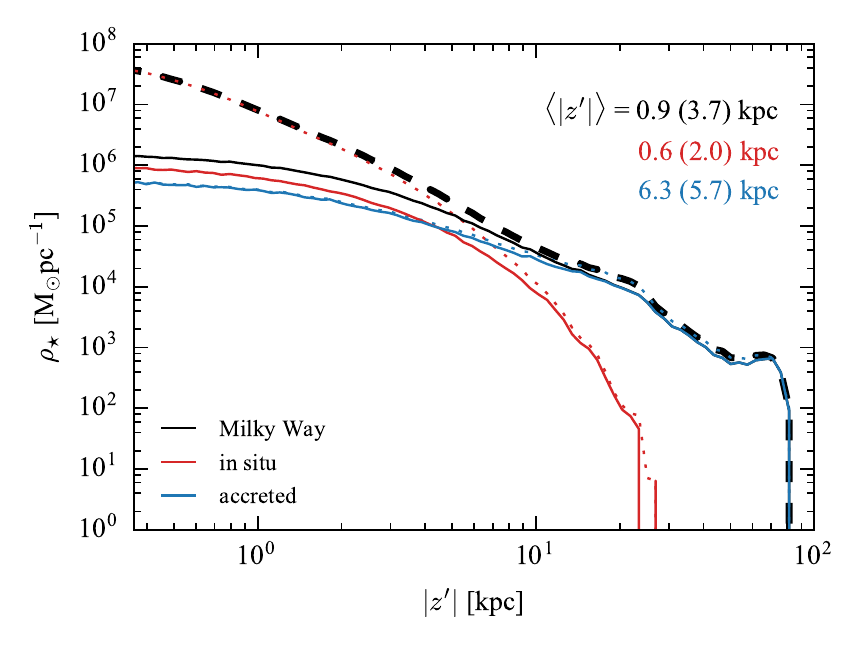}
\caption{Same as \fig {distrib}, but for the radial (in the plane of the disc, top) and vertical (bottom) stellar mass profiles of the Milky Way stars.}
\label{fig:distrib2d}
\end{figure}

\fig[s]{distrib} and \ref{fig:distrib2d} show distributions of the stellar populations of the Milky Way at the end of the simulation. The in situ population is more centrally concentrated and vertically thinner than the accreted one. The accreted population dominates the mass profile beyond $\approx 17 \kpc$ in the plane of the disc and $\approx 7 \kpc$ above. We note that the density profiles are the same for the entire accreted population and for its oldest component only (globular cluster candidates) indicating that early accretion combined with long evolution inside the Milky Way cannot be distinguished from late accretion. This is confirmed by the remarkably flat radial profile of the age of the accreted stars. The globular cluster candidates formed in situ yield comparable distributions than the accreted ones in the inner galaxy ($\lesssim 10 \kpc$), but their fraction drops in the outer parts, making their overall population more centrally (radially and vertically) concentrated than the accreted one.

Star formation being active in the densest regions of the galaxy, the youngest in situ population is mostly found in the central part of the Milky Way. This age gradient is associated with a metallicity gradient of the in situ population, while the accreted one yields very little variation with galactic radius. A comparable behaviour is seen in the vertical profiles (not shown). In the innermost $10 \kpc$ of the simulated galaxy, the metallicity gradient of the Milky Way is $\approx -0.06 \U{dex/kpc}$, in good agreement with observational data \citep[$\approx -0.05 \U{dex/kpc}$,][]{Kewley2010}. In the simulation, the two sub-populations are clearly separated: clusters (globulars and all) formed in situ are more metal-rich and younger than the accreted ones, at all radii in the galaxy. We note that the populations of globular cluster candidates do not yield strong metallicity gradients. However, due to their different spatial distributions, the transition from in situ-dominated in the inner regions to accreted-dominated in the outer areas makes the whole globular population (black solid line in \fig{distrib}) yielding a net metallicity gradient between $\approx 1$ and $20 \kpc$. This qualitatively agrees with the findings of a decreasing [Fe/H] with radial distance in the inner halo and the absence of an [Fe/H] gradient in the outer halo by \citet{Searle1978}. The latter formed a crucial ingredient in their idea on how the Milky Way (halo) grows via (hierarchical) accretion and our model confirms that all globular cluster candidates beyond $\approx 30 \kpc$ are indeed accreted and have similar (average) [Fe/H]. Details on the intensity of this gradient and its radial position depend on how our cluster candidates are representative of real clusters, which is the main uncertainty of our study (see below).

\begin{figure}
\includegraphics{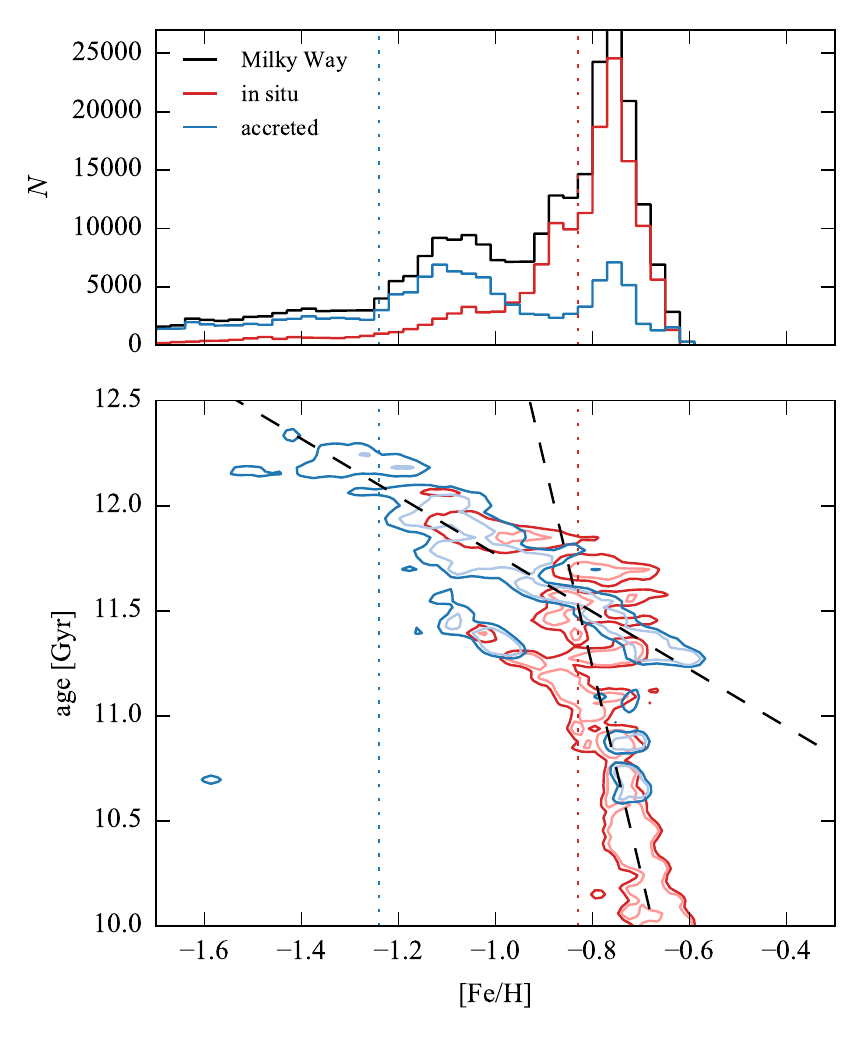}
\caption{Top: distributions of our globular cluster candidates as function of metallicity. Bottom: their locii in the age-metallicity plane. (Contour levels are arbitrary chosen at $10^{-3.3}$, dark colour, and $10^{-3}$, light colour, times the total number of cluster in each sub-population, for the sake of clarity.) Black dashed lines indicate visual fits of the two asymptotic branches (see text) and vertical lines mark average values of the two sub-populations.}
\label{fig:agemetal}
\end{figure}

The top panel of \fig{agemetal} shows the distributions of metallicities for the globular cluster candidates (i.e. older than $10 \Gyr$ at $z=0$). The average values of [Fe/H] = -0.8 and -1.2 for the in situ and accreted populations of globular cluster candidates respectively are close to the values of -0.5 and -1.5 observed for the metal-rich and metal-poor globulars in the real Milky Way \citep{Harris1996}. The Galaxy counts twice more metal-poor ([Fe/H] $< -1$) globular clusters than metal-rich ones \citep{Harris1996}, while in our simulation this ratio is only 0.6. The differences are discussed below.
 
We note that the simulation produces an overall too large number of cluster candidates, but recall that our ``clusters'' are mere stellar particles and that we do not take any cluster evolution processes into account in this study. First, the most massive clusters are likely made of many of our $\sim 10^4 \Msun$ particles, which would thus significantly reduce the number of clusters. Properly inferring the binding energy and thus membership of star clusters would require an unbiased (i.e. non-softened) treatment of gravity at the sub-parsec scale, as shown in \citet{Renaud2015a}. Second, it has been suggested that cluster formation is more efficient in low-metallicity environments than in enriched media \citep{Peebles1984, Fall1985, Bromm2002, Kimm2016}. In this case, the initial cluster mass function would be different for the accreted (i.e. mostly formed in low-metallicity dwarfs) and in situ clusters, making the former more resistant to external disruptions like tides, which could lower the over-abundance of metal-rich objects in our simulation. In addition, we note that at the spacial and mass resolutions of our simulation, we likely underestimate the number of low-mass galaxies (of virial masses below $\sim 10^9 \Msun$), and thus the number of accreted clusters with respect to those form in situ. Furthermore, the stronger tidal field experienced by in situ candidates with respect to the accreted ones (see next Section) could lead to their preferential disruption or even the dissolution of the fragile, low-density end of their distribution, once these aspects would be accounted for. 

The bottom panel of \fig{agemetal} indicates the locii of our globular cluster candidates in the age-metallicity plane. Despite clusters lying on a rather continuous distribution, we note a clear difference of slope between the two ends of the metallicity distribution, making two branches in this diagram. By fitting them by eye, we find that the most metal-rich cluster candidates (with 74 per cent of them formed in situ) yield an age $\propto -9.8 \U{Gyr/dex}$ [Fe/H], while those with lower metallicities (mainly the accreted clusters) present a much more shallower relation: age $\propto -1.4 \U{Gyr/dex}$ [Fe/H]. 

The transition between the two branches originates from the ability of galaxies to produce \emph{and} retain enriched material. The Milky Way being the most massive member of its ``local group'' already at high redshift ($z \approx 5\mh 6$, see \fig{mergertree}), its escape velocity rapidly becomes higher than that of its satellites, which participates in retaining ejecta from feedback, while dwarf galaxies (which have a lower SFR and thus a slower metal enrichment) rather launch outflows and do not efficiently retain metal-enriched medium. Indeed, effective yields of dwarf galaxies are observed to be much lower than that of higher mass galaxies \citep{Tremonti2004}, indicating efficient removal of metals in outflows. This translates in the age-metallicity diagram as follows.

In the young and low-mass Milky Way, early in situ formation produces a few, relatively metal-poor, objects ([Fe/H] $< -1$, see \citealt{Forbes2011} in the context of elliptical galaxies). Rapidly, the galaxy becomes massive enough to produce and retain metals, and it evolves along the shallow branch toward higher metallicities. Then, at an increased SFR, most of the in situ formation proceeds along the steep branch, i.e. at almost constant metallicity. Furthermore, the accretion of low-mass (pristine) galaxies brings metal-poor clusters to the Milky Way. The galaxy population is thus the sum of the in situ group which are mostly metal-rich, and these accreted metal poor clusters, hence making a bimodal distribution. Finally, any galaxy with a mass similar to that of the Milky Way is likely to share a comparable evolution, and thus to yield its own bimodality. Therefore, when such a galaxy merges with the Milky Way, it brings a bimodal accreted population (thus sharing properties with the ``young halo'' cluster population discussed by \citealt{Mackey2005}). In the merger history of our Milky Way, such an event occurs at the last major merger, at $z\approx 2$ (recall \fig{mergertree}).

In short, in agreement with the semi-analytic model of \citet{Tonini2013}, our work suggests that metal-poor clusters form in low-mass, metal-poor galaxies, while the metal-rich ones originate from massive galaxies able to self-enrich their ISM, i.e. mainly the Milky Way and its most massive progenitor galaxies. A merger of two massive enough galaxies (thus typically at late epoch) induces a mixing of accreted and in situ clusters in the high-metallicity mode. Combining the observed distributions of metallicities to that of other tracers of the in situ and accreted populations like the space distribution (see above) could bring constraints on the merger history of the Galaxy.

As noted above, our derived metallicities are slightly off compared to the observational data of Milky Way globular clusters: on average, our metal-poor population is too metallic and our metal-rich one is too metal poor. In addition, our simulation underestimates the relative number of very low metallicity objects ([Fe/H] $\lesssim -1.6$) compared to the real Milky Way. These discrepancies are likely due to the uncertainties in simulating of the assembly of the galaxy, especially at these high redshifts, the imperfections in our modelling of the enrichment at our working resolution (star formation and feedback recipes), the unresolved initial enrichment by population III stars (recall \sect{method}), and the lack of evolution mechanism for our cluster candidates. Furthermore, the resolution of the simulation does not allow us to capture the formation of low-mass dwarf galaxies, which would contribute to the lowest end of the metallicity distribution. However, we notice comparable trends between our age-metallicity relation (\fig{agemetal}) and that observed by \citet[see their Fig. 2]{Leaman2013}. In particular, in the age range we probe, we note a similar slope for our accreted clusters and their halo clusters. Furthermore, there is a remarkable steepening of this relation for Leaman et al.'s disc clusters (at about an age of $12.5 \Gyr$ and [Fe/H]$\approx -1.3$), which could corresponds to the knee in our distribution of in situ cluster candidates (at an age of $\approx 11.5 \Gyr$ and [Fe/H]$\approx -0.8$). Although the position of the clusters and the features in their distributions are sensibly different in our simulation and in reality, the fact that global trends are retrieved indicates the likelihood of the formation scenario we describe, in particular the fact that the bimodality is already in place at $z=2$ \citep[see also][their Figure 10]{Dotter2011}.

\begin{figure}
\includegraphics{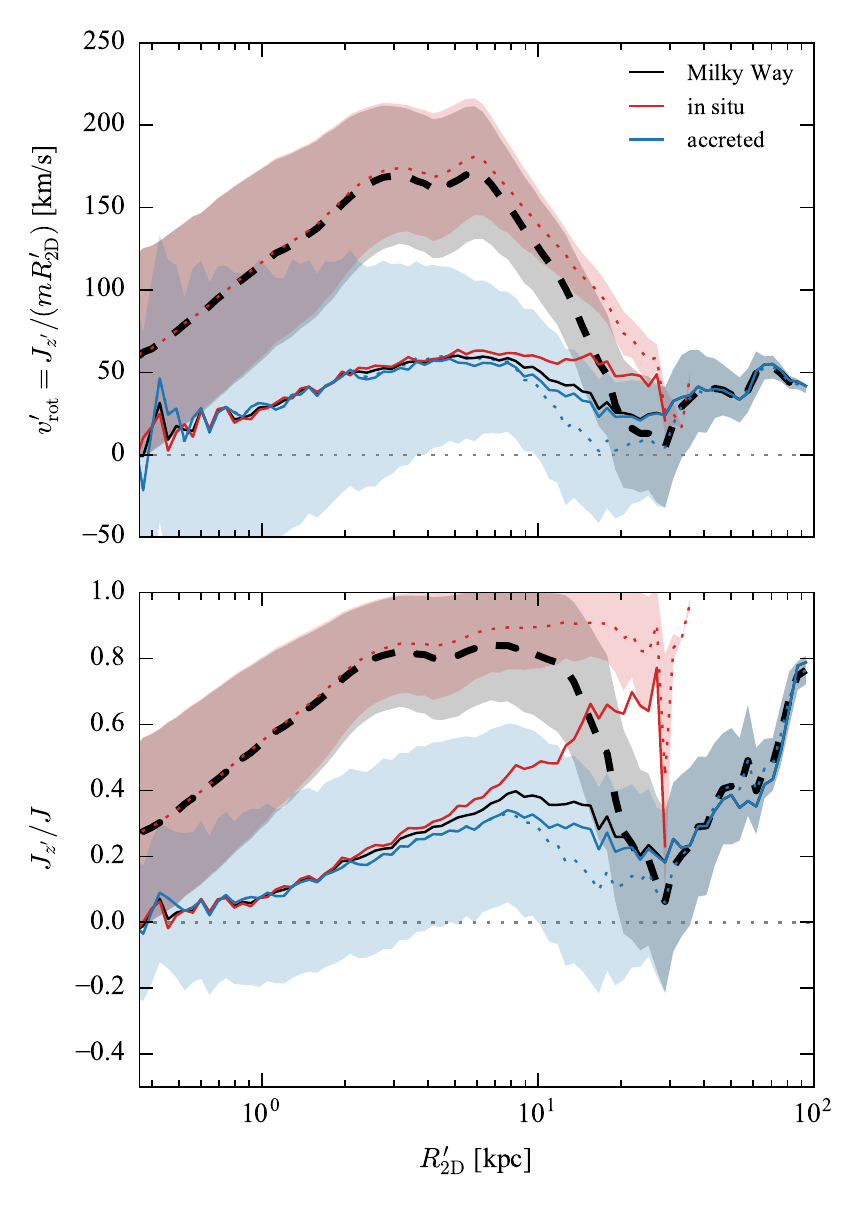}
\caption{Radial profiles of rotation velocity in the plane of the disc (computed from the component of the angular momentum aligned with that of the Milky Way, top) and ratio between the angular momentum aligned along the spin axis of the galaxy $J_{z^{\prime}}$ and its total value $J$ (bottom), for the Milky Way stars (black), those formed in situ (dotted red) and those accreted (dotted blue). Shaded areas indicate the standard deviation in each radial bin. Negative values indicate retrograde motion with respect to the rotation of the disc. As before, solid lines correspond to globular cluster candidates (i.e. older than $10 \Gyr$ at $z=0$).}
\label{fig:angmom}
\end{figure}

Finally, \fig{angmom} highlights the kinematic differences between the in situ and accreted components. Accreted stars retain signatures of their pre-accretion orbital motions for several Gyr. As denoted by the small rotation velocity, the weak alignment with the galactic spin, and the large dispersion of these quantities, these motions are preferentially radial, unlikely to be aligned with the spin of the Milky Way and rather isotropic. Only subtle differences are found between the young and older components (globular clusters) in the outer parts of the galaxy ($R^{\prime}_{2D} \gtrsim 10 \kpc$), where the erasement of the signatures mentioned above is the slowest.

Stars formed in situ have a faster rotation and a spin better aligned with that of the galaxy than the accreted ones. Since the galactic disc is not in place at very high redshift (and is very turbulent and thick in the earliest phases of its evolution), the population of globular clusters formed in situ has kinematic properties closer to that of the accreted globulars than the stars formed more recently in the disc. They however rotate slightly faster and closer to the (final) plane of the disc than the accreted population, especially $\gtrsim 5 \kpc$ from the galactic centre, probably indicating an early influence of the organised dynamics of the galaxy.

\subsection{Tidal histories}

For the volumes considered here, probing the internal physics of the star clusters, or even details of their formation, is still out of reach: the space and time resolutions are not sufficient, and codes treating galaxy evolution lack the physics of clusters like detailed stellar evolution and treatment of the multiple stars. However, by considering our particles as tracers, we can explore the history of the tidal field experienced by our clusters candidates along their orbits. In a forthcoming study, we will complement this analysis with a description of individual clusters and populations of clusters to actually infer their evolution in the environments we consider here.

The formalism and quantities used below are defined in \app{tensor}, where we present a brief reminder about tides. We base our analysis on tidal tensors and their ordered eigenvalues $\lambda_1 \ge \lambda_2 \ge \lambda_3$. For convenience, we normalise the strength of the tides to $\lambdasun$ that we define as the maximum eigenvalue of the inertial tidal tensor of a point of which mass is that enclosed within the orbit of the Sun (considered circular), and evaluated at the position of the Sun in the real Galaxy\footnote{Because of the strong assumptions made, this quantity has no physical meaning. It is however of the order of magnitude of tidal effects for typical orbits in a Milky Way-like galaxy and is thus a convenient scale.}. We use the approximate values for the Local Standard of Rest of $r_{\odot} = 8 \kpc$ and $v_{\odot} = 220 \kms$, which leads us to
\begin{equation}
\lambdasun \equiv \frac{2 G M(< r_{\odot})}{r_{\odot}^3} = \frac{2 v_{\odot}^2}{r_{\odot}^2} \approx 1600 \Gyr^{-2}.
\end{equation}

\begin{figure}
\includegraphics{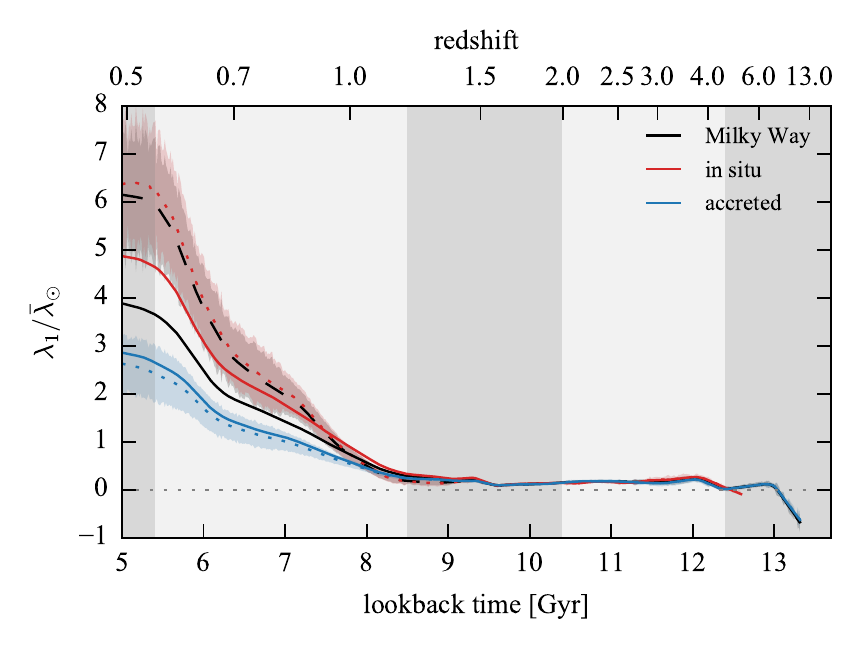}
\caption{Evolution of the instantaneous maximum tidal eigenvalue $\lambda_1$, averaged over members of the sub-populations of particles formed in situ and those being accreted (dotted lines), and only globular clusters (solid lines). This quantity is normalised to $\lambdasun$, which represents a rough proxy of the intensity of the present-day tidal field at the position of the Sun (see text). All curves have been smoothed using a non-parametric local regression method (LOWESS), for the sake of clarity. Shaded areas around the curves represent 0.1 times the standard deviation (non-smoothed). Shaded vertical stripes correspond to the epochs discussed in the text.}
\label{fig:averagel1}
\end{figure}

\begin{figure}
\includegraphics{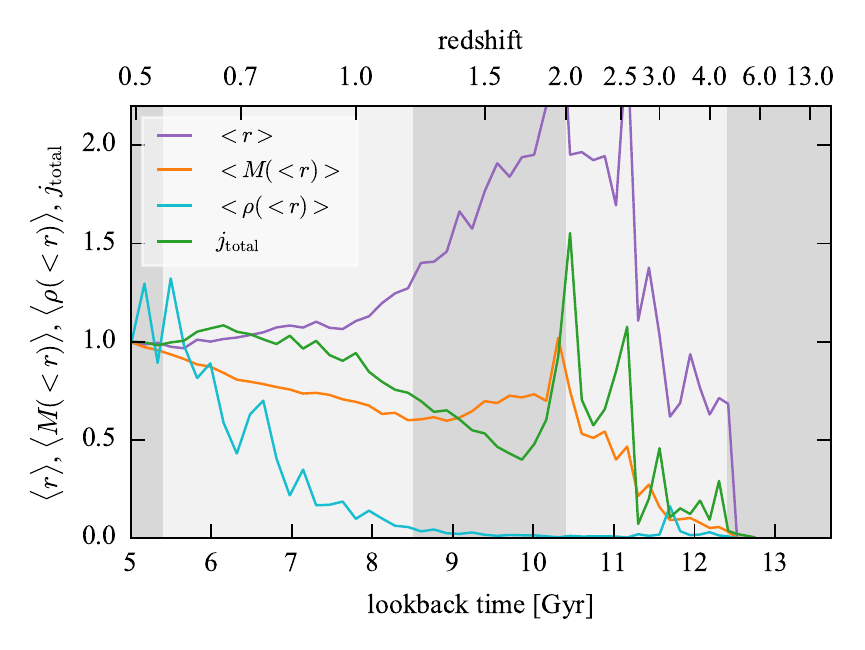}
\caption{Evolution of the averaged radius $\ave{r}$ (purple), the average enclosed mass $\ave{M(<r)}$ (orange), the average enclosed density $\ave{\rho(<r)}$ (cyan), for the in situ stars, and the specific total angular momentum of the Milky Way stars $j_\mathrm{total}$ (green). All quantities are normalized to their final values, respectively $2.93 \kpc$, $3.3 \e{10} \Msun$, $2.8\e{9} \Msun\,\kpc^{-3}$ and $383.0 \U{kpc\,km\, s^{-1}}$. Shaded vertical stripes correspond to the epochs discussed in the text.}
\label{fig:averagemr}
\end{figure}

\fig{averagel1} shows the evolution of the average strength of the tidal field experienced by cluster candidates in our simulation. Reasons for this evolution can be roughly estimated by analysing the average radius $\ave{r}$ of the clusters in the Milky Way, the total mass enclosed in this radius $\ave{M(<r)}$, and the enclosed density $\ave{\rho(<r)} = \ave{3 M(<r) / (4 \pi r^3)}$, as plotted in \fig{averagemr}. (Recall that the tidal field generated by a point-mass galaxy of mass $M$ at a distance $r$ would lead to $\lambda_1 = 2GM/r^3$. See \app{tensor} for details.) In addition, \fig{averagemr} displays the evolution of the specific total angular momentum, computed as the sum of angular momenta for all the stellar particles in the Milky Way divided by the stellar mass of the galaxy, as in \citet{Agertz2016}.

For simplicity, we focus here on the candidate clusters formed in situ. We identify 5 epochs in the evolution:
\begin{enumerate}
\item At high redshift ($z \gtrsim 5$, $\gtrsim 12.4 \Gyr$), the immediate environment of the Milky Way consists of tens of galaxies within a few $10 \kpc$. The superposition of their gravitational potentials (mainly that of their dark matter halos) generates cores spanning several kpc, like those noted during galaxy mergers by \citet{Renaud2009}. This topology corresponds to compressive tides which favour star formation \citep{Renaud2014b}\footnote{At $z\approx 5$, the typical compressive regions in the Milky Way progenitor and its neighbour galaxies span about $5 \kpc$ each, i.e. several tens AMR cells ($\sim 200 \pc$), which ensures that our results are neither affected by the grid nor the softening of the gravitational acceleration. Furthermore the central cusps of the galaxies are also revolved, validating the physical origin of the transition between compressive and extensive tides in our model.}. As the Universe expands and the immediate satellites of the Milky Way get accreted, the number density of galaxies decreases, the Milky Way dominates the gravitational potential, the above-mentioned inter-galactic cores tend to disappear and the tidal fields becomes mostly extensive.
\item During the next $\approx 2 \Gyr$ ($2 \lesssim z \lesssim 5$, between $\approx 10.4$ and $12.4 \Gyr$), the growth of the Milky Way is dominated by discrete but repeated major mergers. The mass of the galaxy rapidly increases, and the transfer of angular momentum leads to a growth in size: both $\ave{r}$ and $\ave{M(<r)}$ increase, resulting in a roughly constant average tidal field strength.
\item Later ($1.2 \lesssim z \lesssim 2$, between $\approx 8.5$ and $10.4 \Gyr$), the SFR reaches its maximum value (\fig{sfr}), with star formation taking place preferentially in the central (dense) regions of the galaxy. The contribution of these newly formed stars reduces the average radius of the ensemble of in situ stars, while $\ave{M(<r)}$ varies little as the result of disc growth (visible as the steady increase of specific total angular momentum, see also \fig{profile}). As a result, the average tidal field strength starts to increase.
\item At $z\approx 1.2$ ($\approx 8.5 \Gyr$), as the disc continues to grow and form stars, the enclosed mass increases while the radius keeps on decreasing. This combine effects translates in the acceleration of the strengthening of tides.
\item At $z \approx 0.6$ ($\approx 5.4 \Gyr$), the decrease in SFR (recall \fig{sfr}) implies that the evolution of both $\ave{r}$ and $\ave{M(<r)}$ slows down. Most of the building up of the Milky Way is done and thus the tidal field is roughly constant. We assume that it would remain constant between the end of the simulated period and the present-day. Local variations might occur in the case of formation of strong substructures like bar(s) and spirals, but we expect the tidal field to remain roughly unchanged when averaged over the entire galaxy.
\end{enumerate}

A similar evolution is expected for the accreted clusters, but only after their accretion. Since this sub-population is made of cluster candidates accreted at different epochs (\fig{sfr}), their average tidal history encompasses both that of the Milky Way as described above (which dominates the statistics of accreted population after the end of the major merger epoch), and that of the satellites galaxies where the clusters formed.

Between the initial phase and the steady increase ($1.2 \lesssim z \lesssim 5$) , the average tidal field of the in situ cluster candidates is comparable to that of the accreted population. This indicates that, before their accretion, clusters experience similar tides whatever their host is. At later times, in situ clusters (all and globular only) become more centrally concentrated in the massive Milky Way, which significantly increases the strength of the tides they experience with respect to their accreted counterparts.

\begin{figure}
\includegraphics{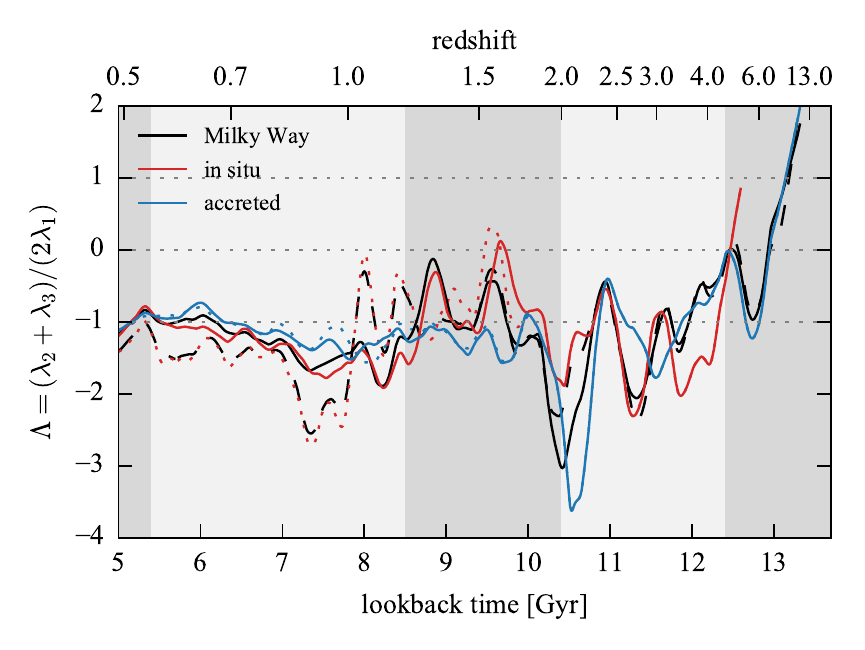}
\caption{Ratio of the average of the minor two eigenvalues of the tidal tensor $\lambda_2$ from $\lambda_3$ to the main one $\lambda1$. Horizontal lines indicates the limits of the tidal regimes (see text for details and interpretation). Shades stripes corresponds to the evolution epochs identified in \fig{averagel1}.}
\label{fig:averagel23}
\end{figure}

To further characterise the nature of the tidal field experienced by the cluster candidates, \fig{averagel23} shows the ratio of the minor eigenvalues to the main one $\lambda_1$. In the non-spherically symmetric configurations we consider, it is not possible to tell apart the physical meaning of $\lambda_2$ from that of $\lambda_3$, and thus we rather consider the average of the two $\lambda_{2,3} = (\lambda_2 + \lambda_3) /2$. Therefore, keeping in mind that the eigenvalues are sorted in descending order, the quantity $\Lambda \equiv \lambda_{2,3}/\lambda_1$ can be:
\begin{itemize}
\item greater than unity (i.e. $0 > \lambda_1 \geq \lambda_{2,3}$), in the case of compressive tides,
\item between 0 and 1 (i.e. $\lambda_1 \geq \lambda_{2,3} \geq 0$), in rare occasions\footnote{Because of Poisson's law, it is impossible that all three eigenvalues are positive (equation~\ref{eqn:poisson}). However, $\lambda_{2,3}$ is positive when only $\lambda_3$ is negative and $\lambda_2 > -\lambda_3$.},
\item between -1 and 0 (i.e. $\lambda_1 > 0 > \lambda_{2,3}$ and $\lambda_1 > -\lambda_{2,3}$), which is the classical case for non-fully compressive tides,
\item smaller than -1 (i.e. $\lambda_1 > 0 > \lambda_{2,3}$ and $\lambda_1 < -\lambda_{2,3}$), which also represent non-fully compressive tides, but with the compression along the minor axes being stronger than the extension along the main axis\footnote{This situation is found e.g. in a spiral arm within a disc: the main tidal axis points toward the galactic centre but the field generated by the arm adds to the compression from the galactic monopole along the perpendicular axes, making it potential stronger then the radial extension.}.
\end{itemize}
The textbook example of the tides generated by a point-mass yields $\Lambda = -1/2$. Smaller (more negative) values can be reached in shallower density profiles. (See also \app{anisotropy} for details.)

In the simulation, the early evolution of $\Lambda$ confirms the compressive nature of tides at the formation phase of the oldest clusters. Quickly, the tidal field jumps to the extensive regime ($\Lambda < 0$). In the major merger dominated phase ($2 \lesssim z \lesssim 5$), $\Lambda$ is mostly smaller than -1, which likely indicates complex substructures in the potential, likely merger-induced, inducing strong transversal compression, in a similar fashion as the effect of spiral arms in a disc (as mentioned above). Later ($1.2 \lesssim z \lesssim 2$), $\Lambda$ rises between -1 and 0 for the in situ population. This corresponds to rather steep density profiles for the Milky Way, that are replaced by shallower ones at the end of the simulation ($z \lesssim 1$), when $\Lambda$ becomes smaller than -1, once the disc has formed (recall \fig{averagemr}). For the accreted population, $\Lambda$ represents a mix of pre- and post-accretion cases and remains close to -1 after the major merger phase.

We note that after $z \approx 1$, the tidal field of any population is highly incompatible with that of the point-mass galaxy ($\Lambda = -1/2$), in contrast with the assumption often made in the literature when modelling Galactic tides.

\begin{figure}
\includegraphics{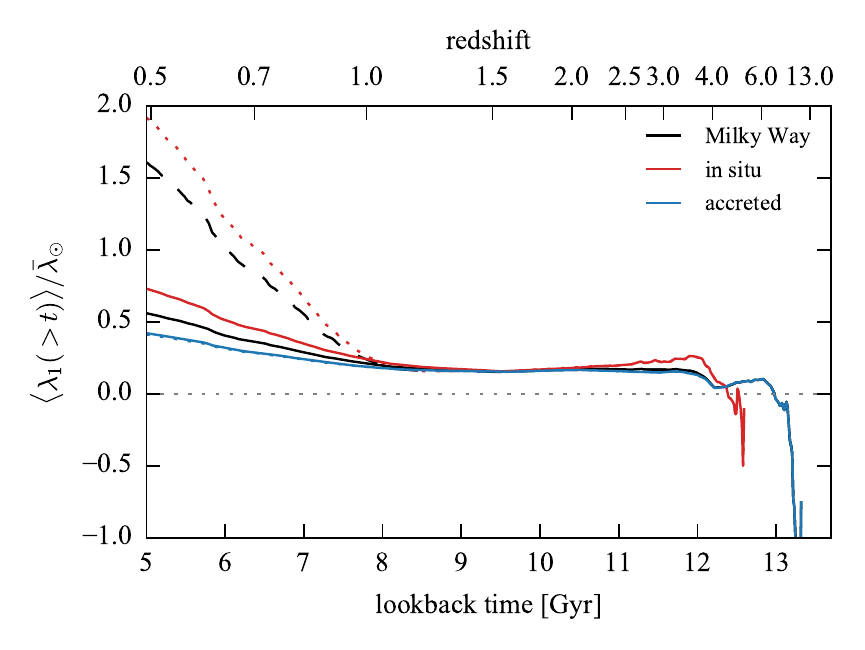}
\caption{Average maximum tidal eigenvalue $\lambda_1$ (normalised to $\lambdasun$, see text) experienced by stars since their birth, i.e. at a given time $t$, the quantity $\ave{\lambda(>t)}$ is the average $\lambda_1$ between the star formation epoch and time $t$.}
\label{fig:avecuml1}
\end{figure}

Finally, \fig{avecuml1} shows, at each time $t$, the value of the average $\lambda_1$ between the formation time and current time $t$. (If, for the purpose of another study, one has to assume a constant tidal field, one should use the final value of this quantity.) The weak tidal field at early stages (mainly due to the low galactic mass) drags down the average of globular cluster candidates, while populations formed later experience stronger fields. Note that both the in situ and accreted globular cluster candidates yield an average smaller than the present-day tidal strength for the Sun ($\lambdasun$).

Furthermore, the present-day tidal field is stronger than the past average, over any period. For in situ cluster candidates, the present-day average tidal field is $\approx 6.3  \lambdasun$ (\fig{averagel1}), while its time-average value ($\approx 1.9 \lambdasun$, over the simulated period) is 3.3 times smaller (\fig{avecuml1}). This ratio jumps to 6.5 for the accreted population. When considering globular cluster candidates only, the present-day values are about 6.8 times stronger than the time-average, for both the in situ and accreted groups, making the present-day value even less representative of the average tides for the globulars than for all clusters candidates. In other words, using the present-day tidal field to model the evolution of clusters over several Gyr, as it is often done, strongly over-estimates the effects of the environment, which significantly affect the conclusions on e.g. cluster mass-loss. When modelling the tidal environment of clusters with a constant field, adopting time-averaged values like those in \fig{avecuml1} would lead to significantly more accurate estimates, as those obtained by the majority of studies. We however argue that a time-dependent environment, with tides varying with time as the cluster itself evolves might lead to yet different and more precise results. 

Using a disruption rate based on the present-day tidal field of the Milky Way, studies argue that the observed peaked cluster mass function could not originate from an initial power-law with index -2 as that observed for young massive clusters \citep[e.g.][]{Baumgardt1998, Vesperini2001, Gieles2008b, Portegies2010}. The strong assumption of time-independent galactic tides makes several author question the validity of this claim, arguing that tides could have a stronger impact on clusters in the earlier Universe \citep{Fall2001, McLaughlin2008b, Kravtsov2005, Muratov2010}. Our results show however that the average tidal field is significantly weaker in the past than at present-day for all cluster populations, and thus that a sole tidal effect cannot explain a depletion of the low-mass end of the cluster mass function and thus a transition from an initial power-law cluster mass function into a peaked one \citep[see also][]{Renaud2015c}. Other mechanisms like the early disruption via encounters with gas clouds might participate in such evolution under extreme conditions (e.g. in gas-rich and clumpy galaxies, see \citealt{Elmegreen2010, Kruijssen2015}, but see also \citealt{Gieles2016}). We will explore this point in more details in a forthcoming paper.

\section{Discussion and conclusions}

We present a cosmological zoom-in simulation of a Milky Way-like galaxy to explore the origins of its star cluster populations. For our study, the most important aspects of the modelled galaxy is the formation of a disc and the absence of major mergers after $z \approx 2$. We characterise the dynamical properties of star cluster populations and their evolution histories. Our main results are as follows.
\begin{itemize}
\item The observed bimodality of globular clusters is reproduced in our model. The physical properties of the modelled sub-populations of globular cluster candidates match those of the observed blue and red groups. Blue, metal-poor clusters originate from low-mass galaxies that are accreted onto the Milky Way. The red, metal-rich clusters form in galaxies massive enough to retain their enriched media, i.e. mostly in the Milky Way itself (in situ clusters), but also in high-mass accreted companions (accreted clusters during late major mergers).
\item The simultaneous formation of the two sub-populations, with differences between them originating from the more rapid growth of the Milky Way than its satellites, implies that all quantities we measure (e.g. space distribution, age, metallicities, tides) yield rather continuous transitions between the two groups, as opposed to the clear separations induced by distinct formation epochs in other scenarios.
\item At high redshift ($z \gtrsim 5$), the high number density of galaxies implies significant overlaps of their gravitational potential wells, making tidally compressive regions over several kpc, in a comparable way as in present-day interacting galaxies. This could affect the star (cluster) formation efficiency.
\item The average tidal field of star cluster candidates remains relatively weak during the early galactic growth phase dominated by major mergers ($z \gtrsim 2$). It strengthens later during the secular build-up through centrally concentrated star formation. 
\item The tidal field experienced by accreted clusters is almost always weaker than that of clusters formed in situ. This is due to their early life in a low-mass (satellite) galaxy and their evolution at large galacto-centric radii (on average) following their accretion onto the Milky Way. Accreted clusters experience tides about four times weaker than those formed in situ (on average over their lifetime).
\item Due to galaxy growth, present-day tidal effects are significantly stronger than their average, over any period in the past, except during the very early phase of galaxy formation when tides are strongly compressive due to the high concentration of galaxies and the overlap of their dark matter halos.
\item Therefore, using the present-day tidal field of the Milky Way to study the dynamical evolution of its star clusters leads to strong over-estimates of their disruption and dissolution rates.
\end{itemize}

These conclusions correspond to the formation history of the galaxy modelled here, but would certainly change when considering other cases. In particular, different merger histories including major mergers at low redshift ($z \lesssim 2$) would significantly alter the distributions of in situ and accreted stars and the relative weight of these sub-populations, as proposed by e.g. \citet{Li2014}. For instance, a recent major merger would \emph{possibly} trigger a starburst activity during which hydrodynamical shocks and compression would lead to the formation of young massive clusters like those observed in the Antennae galaxies \citep[see e.g.][and \citealt{Barnes2004, Renaud2014b, Renaud2015a}]{Whitmore1995, Mengel2005, Bastian2009, Herrera2011}. The merger remnant would then form a massive elliptical galaxy. (Reforming a disc is possible but would take several Gyr, see e.g. \citealt{Athanassoula2016}.) The pre-existing clusters are likely to survive the merger event, as shown in \citet{Renaud2013a}. In this picture, the population of blue globular clusters would form in the progenitor galaxies and would get spatially redistributed by the collision(s). Depending on the orbits and inclinations of the galaxies, a fraction of the pre-existing clusters would be ejected from the remnants (possibly into the tidal debris), and their orbits would span a larger volume after the collision than in the isolated galaxies. On the contrary, metal enriched red clusters would form during the gas-rich merger, in the densest gaseous regions of the merger, and would thus be more centrally concentrated (although starbursting mergers harbour intense star formation activity over several kpc, e.g. \citealt{Wang2004}). This would lead to a cluster bimodality, as suggested by \citet{Ashman1992}, and comparable to the observations of sub-populations in M~87 by \citet[but see the discussion in \citealt{Forbes1997}]{Larsen2001}.

However, this would not explain the bimodality in galaxies which do not experience a recent major merger, like the Milky Way. The scenario we present here allows to form the bimodality with a more quiescent merger history, in line with that of the Milky Way, following \citet{Cote1998}. It is thus likely that the origins of the globular cluster populations vary from one galaxy to the next, being mainly influenced by the galaxy build-up itself. From the hierarchical galaxy formation perspective, a given galaxy would thus encompass the cluster populations of its progenitors, possibly each bringing their own, pre-existing bimodalities. Dynamical mixing during the merger event could then make difficult the identification of sub-populations. For instance, by extending the scenario we propose, a massive elliptical galaxy with mergers throughout its formation history is likely to harbor a rather continuous distribution of accreted clusters, from metal-poor to metal-rich, on top of its in-situ population. Thus, we expect the metallicity distribution to be at least as wide as that of the Milky Way, but with no clear separation between two modes. \citet{Yoon2006} showed that such an uni-modal metallicity distribution could still translate into a bimodal colour distribution due to the non-linearity of the colour-metallicity relations (as observed by, e.g. \citealt{Usher2012}).

The fully compressive nature of the tidal field over large volume at high redshift suggests to draw a parallel between this epoch and the present-day interacting galaxies, where tides become compressive during encounters \citep{Renaud2009}. As noted above, in addition to stopping the destructive effect of classical tides, this mode promotes enhanced star formation, in particular in the form of young massive clusters and is thought of triggering a significant fraction of starburst activity in close galactic pairs \citep{Renaud2014b, Renaud2015a}. Our simulation lacks the resolution to probe the propagation of this tidal effect down to the scale of star forming clouds ($\sim 1\mh 10 \pc$) in the early Universe, but we can speculate that such mechanism does participate in the assembly and collapse of such clouds leading to the formation of massive stellar objects. In that respect, and up to differences due to the extremely low metallicities at high redshift \citep{Fall1985, Kimm2016}, young massive clusters and globular clusters would share similar formation triggers. We note however that their respective evolutions due to early disruption by giant molecular clouds and secular tides would likely differ, implying that currently young massive clusters observed in a few Gyr from now would probably not resemble present-day globulars.

When averaged over the sub-populations, the tidal field experienced by our accreted cluster candidates is significantly weaker than that of the in situ group, partly because of the larger average galactocentric radius of the former. In addition, at a given galactocentric radius in the Milky Way, the time-average tides are also weaker for an accreted cluster, due to its formation in a low mass galaxy. Along those lines, it has been proposed that accretion onto the Milky Way could explain the observed extended size of accreted globulars, compared to those formed in situ \citep{Mackey2004}. This question has been investigated using $N$-body star-by-star simulations of clusters accounting for both the internal evolution and a time-dependent tidal field mimicking the accretion onto the Milky Way. Theses studies show that clusters rapidly adjust to their new tidal environment once they are accreted onto the Milky Way, such that no distinction between the in situ and accreted population could be made based on their size only \citep{Bianchini2015, Miholics2016}. Therefore, the observed larger size of accreted objects is more likely a signature of their formation and/or early evolution, rather than a result of their accretion event itself \citep{Elmegreen2008}.

Because of resolution limitations, this work misses a detailed description of the properties of the clusters at the epoch of their formation. Modelling the mass and size of star clusters implies a proper treatment of turbulence and cooling mechanisms within their formation sites, which requires at least parsec resolution. (See the non-convergence of star formation rates when the turbulence cascade is not resolved at least up to the supersonic scale, as illustrated in Fig. 1 of \citealt{Renaud2014b}. See also \citealt{Teyssier2010}). Such resolutions are now commonly reached in galaxies simulations neglecting the cosmological context, and start to be in range of full cosmological setup, as demonstrated by \citet[which is however limited to high redshifts]{Li2016}. 

Furthermore, all the clusters we consider are mere candidates since we do not model their mass-loss and potential dissolution. Our results are thus biased towards over-estimating the contribution of clusters that would get disrupted early, like low-density clusters and those in strong tidal fields. However, it is likely that typical tides would only have moderate effects on the dense, resistant globular clusters. Using the tidal information extracted from this simulation, we will study the response of individual clusters and entire populations to these tidal perturbations in a forthcoming contribution.

\section*{Acknowledgements}
We thank Phil Hopkins and Andrew Wetzel for sharing the \music parameter file for generating the initial conditions used in this work. We thank Duncan Forbes, Oleg Gnedin, Justin Read and Sukyoung Yi for discussions which helped to improve this paper, and the anonymous referee for a constructive report. This work was granted access to the PRACE Research Infrastructure resource \emph{Curie} hosted at the TGCC (France). FR and MG acknowledge support from the European Research Council through grant ERC-StG-335936. OA would like to acknowledge support from STFC consolidated grant ST/M000990/1 and the Swedish Research Council (grant 2014-5791). MG acknowledges financial support from the Royal Society (University Research Fellowship).

\bibliographystyle{mnras}
\bibliography{biblio}

\appendix

\section{A brief reminder on tides, tidal tensors and tidal radii}
\label{sec:tensor}

Galactic tides are treated from many different perspectives in the literature, which is causing confusion, especially on the validity range of formulae and the equivalence of various formalisms. Here, we attempt a clarification, starting from the most general formalism offered by tidal tensors.

At any given point in space and time, the tidal field can be fully and exactly described by the tidal tensor, which is minus the Hessian matrix of the gravitational potential $\phi$. Its component for the $i$-th and $j$-th space coordinates reads
\begin{equation}
T^{ij} = -\frac{\partial^2 \phi}{\partial x_i \partial x_j}.
\end{equation}

\subsection{Eigenvalues of the tidal tensor}

It is often convenient to write a tidal tensor in its diagonal form, where the three eigenvalues $\lambda_1 \geq \lambda_2 \geq \lambda_3$ represent the intensity of the tidal field along the corresponding eigenvectors\footnote{Note that the base formed by the eigenvectors makes, in general, a non-inertial reference frame. This implies that reconstructing the tidal effect in such a frame requires to include fictitious terms, in the form of the centrifugal, Coriolis and Euler forces \citep[see][for more details]{Renaud2011}. These additional terms would then represent the acceleration of the eigenbase in an inertial reference frame.}.

In spherically symmetric potentials, the eigenvector associated with the maximum eigenvalue points toward the centre of potential, and the other two eigenvalues are strictly equal. In a classical situation like tides generated by a point-mass (e.g. the Moon on the Earth), the maximum eigenvalue is positive and the other two are negative.

It is useful to note that the trace of the tidal tensor, i.e. the sum of its eigenvalues, is minus the Laplacian of the gravitational potential and thus, from Poisson's law, one has
\begin{equation}
\label{eqn:poisson}
{\rm Tr}(T) = -\left(\frac{\partial^2 \phi}{\partial x^2}+\frac{\partial^2 \phi}{\partial y^2}+\frac{\partial^2 \phi}{\partial z^2}\right) = - \nabla^2 \phi = -4\pi G \rho \leq 0
\end{equation}
where $\rho$ is the local density. Therefore, it is impossible to have three strictly positive eigenvalues.

Negative eigenvalues represent tidal compression along the corresponding axis. When all eigenvalues are negative (i.e. the tensor is negative definite), the potential is convex (or cored) and the tidal field is fully compressive \citep[see e.g.][]{Renaud2008, Renaud2009}.

\subsection{Inertial reference frame and tidal radius}

For a star cluster of mass $m$ embedded in a tidal field, the \emph{inertial} tidal radius is
\begin{equation}
\label{eqn:rt}
r_\mathrm{t} = \left(\frac{Gm}{\lambda_1}\right)^{1/3}.
\end{equation}
(Note that obtaining this expression, and all the following, requires to linearise the tidal field at the position of the cluster, and to assume the potential of the cluster itself can be considered as that of a point-mass at the distance $r_\mathrm{t}$ from its centre.) Using this definition for a cluster at a distance $R$ from a point-mass galaxy of mass $M$, one gets 
\begin{equation}
\label{eqn:rtpt}
r_\mathrm{t, point} = R \left(\frac{m}{2M}\right)^{1/3},
\end{equation}
as in e.g. \citet[his equation 33, see also \citealt{Spitzer1987}]{vonHoerner1957}.

The tidal radius is however often written as
\begin{equation}
\label{eqn:rtfictitious}
r_\mathrm{t,e} = \left(\frac{Gm}{\Omega^2 - \frac{\partial^2 \phi}{\partial R^2}}\right)^{1/3},
\end{equation}
using the angular velocity $\Omega$ \citep{King1962}, which implies that the expression for a point-mass potential is
\begin{equation}
\label{eqn:rtfictitiouspt}
r_\mathrm{t,e, point} = R\left(\frac{m}{3M}\right)^{1/3} = R\left(\frac{Gm}{3\Omega^2}\right)^{1/3},
\end{equation}
which is different from \eqn{rtpt} by a factor $(2/3)^{1/3} \approx 0.87$.

The difference between the \eqn[s]{rt} and (\ref{eqn:rtfictitious}) is that no assumption is made on the cluster orbit in the former case (which thus does not include non-inertial terms), while the latter is a specific case including the centrifugal force along a circular orbit. This centrifugal effect is represented by the $\Omega^2$ term in \eqn{rtfictitious}. This can also be expressed in term of eigenvalues of the inertial tidal tensor by noting that, for any circular orbit in a spherically symmetric potential, $\Omega^2 = -\lambda_2 = -\lambda_3$. Thus, \eqn{rtfictitious} is strictly equivalent to
\begin{equation}
\label{eqn:rtfictitiouslambda}
r_\mathrm{t,e} = \left(\frac{Gm}{\lambda_1 - \lambda_2}\right)^{1/3}.
\end{equation}
Note that $\lambda_1 - \lambda_2$ is called the \emph{effective} eigenvalue, noted $\lambda_{\mathrm{e},1}$ in \citet{Renaud2011} \emph{for circular orbits in spherically symmetric potentials only}.

In other words, \eqn{rt} represents the tidal radius in an inertial reference frame, while \eqn[s]{rtfictitious} and (\ref{eqn:rtfictitiouslambda}) correspond to a frame co-rotating with the cluster, along a circular orbit. (Because of the complexity introduced by Coriolis terms, there is no analytical expression for an effective tidal radius along an elliptical orbit, but see an approach based on perturbation theory by Bar-Or et al., in preparation, and its numerical counterpart in \citealt{Cai2016}, following \citealt{Baumgardt2003}.)

Since both expressions for the tidal radius are valid for clusters on circular orbits, which should be considered when studying, e.g., the mass-loss of clusters in a tidal field? None. In such problem, the orbital motion of individual stars in the cluster should be considered. For instance, prograde orbits (i.e. the angular momentum of the star within the cluster is aligned with the angular momentum of the cluster in the galaxy) yield a stronger effective centrifugal effect, which allows stars on these orbits to escape the cluster with a lower energy than those on retrograde orbits \citep[see e.g.][]{Read2006, Tiongco2016b}. In other words defining a global tidal radius (or equivalently, an escape energy level) for the cluster cannot correctly describe the actual escape of cluster members. The tidal radius (or the tidal energy at the Lagrange point L$_1$) is a global quantity defined by considering the cluster as a single object. Thus, it cannot be used to assess the properties of some constituents of this object, like the escape rate. Such shortcut is one of the reasons for the existence of potential escapers \citep[i.e. stars with an energy above that of the Lagrange point L$_1$, but still within the Jacobi surface for several orbits, see][Claydon et al., submitted]{Henon1969, Fukushige2000, Baumgardt2001}.

\subsection{Anisotropy of the tidal field}
\label{sec:anisotropy}

Because the three eigenvalues can never be all strictly positive (equation~\ref{eqn:poisson}), an isotropic tidal field can only be found in fully compressive tides ($\lambda_1 = \lambda_2 = \lambda_3 < 0$), or a tide-free environment ($\lambda_1 = \lambda_2 = \lambda_3 = 0$).  In all other cases, the ratios of eigenvalues define the anisotropy of the tidal field, which can be pictured using the flattening of the Jacobi surface \citep[see e.g.][their Figure 1 and equation 14]{Renaud2011}. 

\begin{figure}
\includegraphics{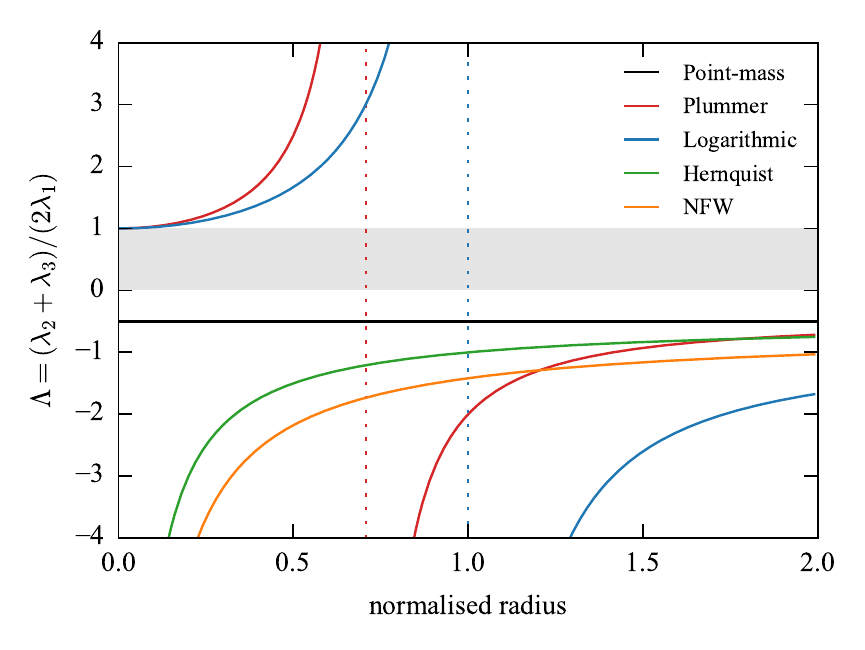}
\caption{Ratio $\Lambda = (\lambda_2 + \lambda_3) / (2\lambda_1)$ for several analytical density profiles. The radius is normalised to the scale-radius of the profile \citep[see Appendix B of][for details]{Renaud2010a}. Vertical dotted lines are the asymptotes marking the transition from compressive to extensive regime for the Plummer and logarithmic profiles. The shaded area indicates the ``forbidden'' regime, corresponding to negative densities.}
\label{fig:analyticl23}
\end{figure}

\fig{analyticl23} shows the ratio $\Lambda = (\lambda_2 + \lambda_3) / (2 \lambda_1)$ derived analytically for some classical density profiles (which is exactly $= \lambda_2/\lambda_1 =\lambda_3/\lambda_1$ for these spherically symmetric profiles). For a given $\lambda_1$, different $\Lambda$'s would have different effects on the object embedded in the tidal field. 

For instance, \citet{Tanikawa2010} show that, for star clusters orbiting in a power-law density profile and \emph{for a given tidal radius}, the shallower the profile, the slower the mass-loss. In term of eigenvalues, this statement translates into: for a given $\lambda_1$, the smaller $\Lambda$ (i.e. negative with a large absolute value), the slower the mass-loss. Again, this can be pictured using the Jacobi surface. By setting the tidal radius, one fixes the position of the Lagrange point L$_1$. Varying the slope of the density profile (or equivalently the ratio $\Lambda$) produces different flattening of the surface: a shallow density profile (i.e. a small, very negative, $\Lambda$) corresponds to a flattened surface and a smaller aperture around L$_1$ and L$_2$ in the equipotential surface for the stars to escape through (for a given energy above that of L$_1$). This is illustrated in figure 3 of \citet{Renaud2011}. Following \citet{Fukushige2000}, and in the special case of circular orbits, \citet{Renaud2011} estimate the dissolution time of a cluster (their equation 29) to go as
\begin{align}
t_\mathrm{diss} & \propto  r_\mathrm{t,e}^{3/2} \left( 1-\frac{\lambda_{\mathrm{e},3}}{\lambda_{\mathrm{e},1}} \right)^{1/8} \nonumber \\
& \propto r_\mathrm{t,e}^{3/2} \left( 2+\frac{1}{\Lambda -1} \right)^{1/8},
\end{align}
confirming that shallower galactic profiles induce longer dissolution times (see also Claydon et al., submitted).


\end{document}